\documentclass[preprints,article,accept,pdftex,moreauthors]{Definitions/mdpi} 
\firstpage{1}
\pubvolume{7}
\issuenum{3}
\articlenumber{30}
\pubyear{2025}
\copyrightyear{2025}
\datereceived{22 April 2025}
\daterevised{27 June 2025} % Comment out if no revised date
\dateaccepted{8 July 2025}
\datepublished{22 July 2025}
\hreflink{https://doi.org/10.3390/\linebreak  physics7030030} % If needed use \linebreak
\usepackage{xspace,nicefrac,bm}
\def\TeV{\ifmmode {\mathrm{Te\kern -0.1em V}}\else
\textrm{Te\kern -0.1em V}\fi}%
\def\GeV{\ifmmode {\mathrm{Ge\kern -0.1em V}}\else
\textrm{Ge\kern -0.1em V}\fi}%
\def\MeV{\ifmmode {\mathrm{Me\kern -0.1em V}}\else
\textrm{Me\kern -0.1em V}\fi}%
\def\keV{\ifmmode {\mathrm{ke\kern -0.1em V}}\else
\textrm{ke\kern -0.1em V}\fi}%
\def\eV{\ifmmode  {\mathrm{e\kern -0.1em V}}\else
\textrm{e\kern -0.1em V}\fi}%
\let\tev=\TeV
\let\gev=\GeV

\def\Pythia{\textsc{Pythia8}\xspace}
\def\Herwig{\textsc{Herwig 7.3}\xspace}

\def\ifb{\mbox{fb$^{-1}$}\xspace}%  Inverse femtobarns.
\newcommand{\eee}{\ensuremath{e^+e^-}\xspace}
\newcommand{\half}{\nicefrac{1}{2}\xspace}
\Title{Exploring Hidden Sectors with Two-Particle Angular Correlations at Future $e^{+}e^{-}$ Colliders}

\TitleCitation{Exploring Hidden Sectors with Two-Particle Angular Correlations at Future $e^{+}e^{-}$ Colliders}

\Author{{Emanuela~Musumeci} 
$^{1}$\href{https://orcid.org/0000-0002-2077-4949}{\orcidicon}, Adri\'an~Irles $^{1,}$*\href{https://orcid.org/0000-0001-5668-151X}{\orcidicon}, Redamy~P\'erez-Ramos $^{2,3}$\href{https://orcid.org/0000-0002-4243-0063}{\orcidicon}, Imanol~Corredoira $^{4}$\href{https://orcid.org/0000-0002-6089-0899}{\orcidicon}, Edward~Sarkisyan-Grinbaum $^{5,6}$\href{https://orcid.org/0000-0003-1794-5871}{\orcidicon}, Vasiliki~A.~Mitsou $^{1}$\href{https://orcid.org/0000-0002-1533-8886}{\orcidicon} and {Miguel~\'Angel~Sanchis-Lozano} 
$^{1}$\href{https://orcid.org/0000-0003-2666-7711}{\orcidicon}}

\AuthorNames{Emanuela~Musumeci, Adri\'an~Irles, Redamy~P\'erez-Ramos, Imanol~Corredoira, Edward~Sarkisyan-Grinbaum, Vasiliki~A.~Mitsou and Miguel~\'Angel~Sanchis-Lozano}

\isAPAStyle{%
\AuthorCitation{Musumeci, E., Irles, A., P\'erez-Ramos, R., Corredoira, I., Sarkisyan-Grinbaum, E., Mitsou, V.A., \& Sanchis-Lozano, M.A.}
}{%
\isChicagoStyle{%
\AuthorCitation{Musumeci Emanuela, Irles Adri\'an, P\'erez-Ramos Redamy, Corredoira Imanol, Sarkisyan-Grinbaum Edward, Mitsou Vasiliki A., and Sanchis-Lozano Miguel~\'Angel.}
}{
\AuthorCitation{Musumeci, E.; Irles, A.; P\'erez-Ramos, R.; Corredoira, I.; Sarkisyan-Grinbaum, E.; Mitsou, V.A.; Sanchis-Lozano, M.A.}
}
}

\address{$^{1}$ \quad {IFIC (Instituto de F\'{\i}sica Corpuscular), %MDPI: No undefined abbreviations, variables, quantities, indexes etc. as met first in each of the main parts of a paper such as afiliations, abstract, mainbody text, figures, tables even known. Please consider. Here and elsewhere below.
Universitat} %MDPI: The department/school/faculty/campus of this university is required. Please try to provide this information.
de Val\`encia and CSIC {(Consejo Superior de Investigaciones Cient\'{\i}ficas)},
46980 Paterna, Spain; emanuela.musumeci@ific.uv.es (E.M.); vasiliki.mitsou@ific.uv.es (V.A.M.);
mas@ific.uv.es (M.\'A.S.-L.)\\ %Author: there is no departament or faculty. The correct affiliation is what it is listed now. %Author: I added adrian.irles@ific.uv.es mail to the list
$^{2}$ \quad {DRII}--IPSA {(Direction de la Recherche et de l'Innovation de l'Institut Polytechnique des Sciences Avanc\'ees)}%MDPI: Please confirm the full name of this affiliation; same just below %Author: corrected by removing "Disaster Recovery..."
, 94200 Ivry-sur-Seine, France; redamy.perez-ramos@ipsa.fr\\
$^{3}$ \quad Laboratoire de Physique Th\'eorique et Hautes Energies (LPTHE), UMR 7589, Sorbonne Universit\'e et
Centre National de la Recherche Scientifique  (CNRS), 75252 Paris Cedex 05, France \\
$^{4}$ \quad Instituto Galego de F\'isica de Altas Enerx\`ias (IGFAE), Universidade de Santiago de Compostela, \linebreak 15782 Santiago de Compostela, Spain; imanol.corredoira.fernandez@cern.ch\\
$^{5}$ \quad Experimental Physics Department, {CERN (European Organization for Nuclear Research)},  {1211 Geneva 23,}
Switzerland; edward.sarkisyan-grinbaum@cern.ch\\
$^{6}$ \quad Department of Physics, The University of Texas at Arlington, Arlington, TX 76019, USA
}

\corres{Correspondence: adrian.irles@ific.uv.es}

\abstract{
Future $e^+e^-$ colliders are expected to play a fundamental role in measuring Standard Model (SM) parameters with unprecedented
precision and in probing physics beyond the SM (BSM). This study investigates two-particle angular correlation distributions
involving final-state SM charged hadrons. Unexpected correlation structures in these distributions
{is considered to } %MDPI; fits better;; mismatching "can", "could" recommended to be avoided. Please consider. %Author: accepted
be a hint for new physics perturbing the {QCD} %MDPI: not defined. As many abbreviations used, all are summarised in the Abbreviations section end of the paper. Same for all other (defined or not) abbreviations. Please consider.  %Author: we checked the list, we agree with the list.
partonic cascade and thereby modifying azimuthal and (pseudo)rapidity
correlations. Using \textsc{Pythia8}
{Monte Carlo generator} %MDPI: no jargon even understood in the field. Here and elsewhere below.   %Author: accepted
and fast simulation, including selection cuts and detector effects, we study potential structures in the two-particle angular
correlation function. We adopt the QCD-like Hidden Valley (HV)
scenario as implemented in  \textsc{Pythia8} {generator}, with relatively light {HV} %MDPI: this explains the "v" prefix. Please consider, otherwise "v" is very unclear. %Author: accepted
$v$-quarks ({below about} %$\lesssim$ %MDPI: no symbols within the text unless a full equation given. Please consider. Here and elsewhere below.  %Author: accepted
100~GeV), to illustrate the potential of \mbox{this method.}}

\keyword{$e^+e^-$ colliders; angular correlations; ridge {phenomenon}; %MDPI: please confirm, as "phenomena" is better suitable.
new physics; Hidden Valley; Higgs factories}

\begin{document}
%%%%%%%%%%%%%%%%%%%%%%%%%%%%%%%%%%%%%%%%%%
\section{Introduction}
\label{sec:intro}

Correlations play a fundamental role in the study of hadronic dynamics since the beginning of cosmic ray and accelerator
physics~\cite{Kittel:2005fu,Botet:2002gj}. Recently, the study of angular correlations {\color{black} has} revealed a new
phenomenon in heavy-ion collisions at both the BNL RHIC~\cite{STAR:2005ryu,PHOBOS:2008yxa,PHOBOS:2009sau,STAR:2009ngv} and the CERN
LHC~\cite{ALICE:2010suc,CMS:2011cqy,CMS:2012xss,ATLAS:2012at,CMS:2013bza}, later also found in smaller system
(\mbox{proton--proton}, proton--nucleus)
collisions~\cite{CMS:2010ifv,CMS:2012qk,ALICE:2012eyl,ATLAS:2012cix,CMS:2013jlh,ALICE:2013snk,CMS:2014und,CMS:2015yux,ATLAS:2015hzw,CMS:2015fgy,LHCb:2015coe,CMS:2016fnw,ATLAS:2017hap,ATLAS:2017rtr,ATLAS:2023bmp,CMS:2023iam,CMS:2025zhz}.

In particular, long-range near-side (so-called) ridges appear in two-particle angular correlations, while different theoretical
explanations have been put forward to
understand this initially unexpected phenomenon
\cite{Shuryak:2007fu,Dumitru:2008wn,Bozek:2013uha,Dusling:2015gta,Sanchis-Lozano:2016qda,Sanchis-Lozano:2017aur,Noronha:2024dtq}.
Almost all of {those theories}  %them %MDPI: The full-wording instead of "it", "them", "their" etc. recommended as the latter are mismatching, ambiguous, making text very unclear. Please consider. Here and elsewhere below. %Author: accepted
require the existence of {an} %some  %Author: accepted
unconventional state of matter at the primary interaction of the collision (like{, for example, } %MDPI; thsi si not an only state as it should be.
quark--gluon plasma), ultimately yielding collective effects among final-state SM particles.
On the other hand, no %clear %MDPI: No "clear/ly", "undoubted/ly", "evident/ly", "of course", "obviously" etc. recommended for scientific papers. Please consider replacing or removal. Here and elsewhere below. %Author: accepted explicit
{explicit}
ridge-like signal was found in the \eee
data analysed by ALEPH~\cite{Badea:2019vey} and Belle~\cite{Belle:2022fvl,Belle:2022ars}
experiments, except for
recent claims in Ref.~\cite{Chen:2023njr} at the highest available energy and high multiplicity of the former.

%Motivated by the unexpected correlation structures observed in high-energy collisions, we investigate possible anomalies—not
%limited to ridge effects—in both azimuthal and (pseudo)rapidity correlations, aiming to gain insights into unknown stages of matter
%beyond the standard QCD parton shower. We focus on two-particle angular correlations as a tool to search for BSM phenomena at
%future high-energy colliders, which offer a significantly cleaner environment than hadron colliders in both theoretical and
%experimental respects. For example, QCD effects associated with matter under extreme temperature and density conditions—such as
%those leading to the formation of a Gluon Saturation state or the {color glass condensate} (CGC) \cite{Gelis:2010nm}—cannot be
%reproduced in collisions due to the absence of initial-state gluonic fields and beam remnants. Finally, we emphasize that the scope
%of this study is limited to the search for new phenomena in leptonic collisions, and does not extend to hadron colliders such as
%the LHC.

{Motivated} %MDPI: we have removed the previous paragraph which (with some tiny differences) repeats this paragraph. Please confirm.  %Author: accepted
by such unexpected correlation structures found in high-energy collisions, we {\color{black} investigate} possible
anomalies (not {\color{black} limited to} ridge effects) in both azimuthal and (pseudo) rapidity
correlations { %aiming %%MDPI: No the future tenses ("will", "shall", "would" etc.), aiming, promising for the current paper, findings, statements etc. Please consider. here and elsewhere below. %Author: accepted
to clarify on %gain %MDPI: colloqual, not recommended fore scientific papers. Please consider. Here and elsewhere below.
%a  %Author: accepted
insights} into unknown stages of matter on top of the QCD parton shower. We focus on
two-particle angular correlations as a way of searching for BSM at future high-energy $e^+ e^-$ colliders, {\color{black} which
offer} a {\color{black} significantly} cleaner environment than hadron colliders in {\color{black} both} theoretical and
experimental respects. For instance, QCD effects on matter under extreme temperature and density conditions
({for example}%e.g. %MDPI: "for example", "for instance" etc. full-wording for the text, while "e.g." for sets of numbers, refs etc. numbering. Please consider. %Author: accepted
, leading, to the
%Glass--Gluon Condensate
{gluon--gluon condensate}
%MDPI: Please check intended meaning has been retained. %Author: accepted
\cite{Gelis:2010nm}) cannot be reached in $e^+ e^-$ collisions due to the {\color{black} absence} of {\color{black} initial-state}
gluonic fields and beam remnants. Let us also emphasize that the scope of the {\color{black} this} study {\color{black}is limited}
to the search for new phenomena in leptonic collisions, and {\color{black} does not trivially extend to} hadron colliders
{\color{black} such as} the LHC.

The theoretical framework of {new physics} %MDPI: no capitlization necessary. Please consider. %Author: accepted
(NP) {\color{black} explored} in this
%work
{paper} %MDPI: The "paper", "study", "investigation", "review" etc. recommended instead "work", "article", "analysis", "manuscript", "reference" for scientific papers. Please consider. Here and elsewhere below. %Author: accepted
is based on the so-called Hidden Valley (HV) scenario, which actually {\color{black} encompasses a broad} class of models with one or more hidden sectors beyond the SM. Thereby, new (valley) particles {\color{black} with masses}
{below about} %$\lesssim$
100 GeV {\color{black} can} arise, surviving {\color{black} current experimental constraints (see discussion in \mbox{Section \ref{sec:hv_pheno})}, and contribute} to the partonic cascade, {\color{black} ultimately} yielding final-state \mbox{SM particles.}

\section{Two-Particle Angular Correlations}
\label{sec:2pac}

%Note that only rapidity and azimuthal differences between two final-state SM charged particles—labelled as ‘1’ and ‘2’—are considered.

{The} %MDPI: we removed teh previous paragraph as it is useless given being repeated and extended at the end of this paragraph. Also it is not at all recomemnded to start a paragraph with introducing the notatrions, moreover not used so far, very unclear, mismatching. Please consider.  %Author: accepted
clean {enough}
environment of \eee collisions, in contrast to hadronic collisions, is {\color{black} particularly well-suited for defining}
a reference frame whose $z$-axis  {\color{black} aligns with} the direction of the back-to-back jets in most events.
{%In this work we adopt the
{The} thrust reference frame} {is adopted in this study, }
so the rapidity $y$ of a particle {\color{black} is always defined with respect} to the thrust or $z$-axis. The azimuthal angle
$\phi$ is defined, as %usual %MDPI: ("usual/ly") colloquial, not recommended for scientific papers. Please consider. Here and elsewhere below. %Author: accepted
{most ofetn considered,}
on the transverse plane to the thrust axis, {\color{black} on an
event{-by-}event %MDPI: dashes necessarily added. %Author: accepted
basis}.
Note that only rapidity and azimuthal{-angle (azimuthal)} %MDPI: added as clarifying "azimutthal" used in what comes below. Please consider. %Author: accepted
differences of two final-state SM charged {particles %labelled as %`1' and `2',
1 and 2,} %MDPI: no any quotation necessary.  %Author: accepted
$\Delta y \equiv
y_1-y_2$ %,
{and} $\Delta \phi \equiv \phi_1-\phi_2$,
{respectively,} %MDPI; otherwise unclear correspondence. Please consider.   %Author: accepted
{\color{black} are considered } %in
%our %MDPI:No "we", "our" recommended for scientific papers but "the", "this", "current", otherwise highly mismatching with possible other studies beyong the current one, no varaibles, indexes, formulas etc. to be considered as "our" as not belongings to the authors, etc.  Please consider throughout. %Author: accepted
{throughout the }
study.

The two-particle correlation function is defined as
%the ratio
\begin{equation}
C^{(2)}(\Delta y, \Delta \phi) = \frac{S(\Delta y, \Delta \phi)}{B(\Delta y, \Delta \phi)},
\label{eq:corr_function}
\end{equation}
where $S(\Delta y, \Delta \phi)$ stands for the density of particle pairs within the same event{: }
\begin{equation}
S(\Delta y, \Delta \phi) = \frac{1}{N_\text{pairs}} \frac{d^{2}N_\text{same}}{d \Delta y d \Delta \phi},
\label{eq:same_evt}
\end{equation}
while $B(\Delta y, \Delta \phi) $ represents the density of
mixed particle pairs from distinct {events:} %MDPI: we have added "pairs" to the N_mix in the denominators to make eqs(2) and (3) similarly reading. Please confirm. Otherwise looks quite mismatching. %Author: accepted

\begin{equation}
B(\Delta y, \Delta \phi) = \frac{1}{N_\text{pairs,\, mix}} \frac{d^{2}N_\text{mix}}{d \Delta y d \Delta \phi}.
\label{eq:mixed_evt}
\end{equation}
{Here, $N_{\rm same}$  denotes the number of events where the pairs of particle of the total number $N_{\rm pairs}$ of all events
were selected per same event, and
$N_{\rm mix}$ denotes the number of events from where the pairs of a total number $N_{\rm pairs,\,  mix}$ of pairs with each
particle per a pair from different events for all such events were selected.}  %MDPI: please confirm, otherwise stands very unclear.   %Author: accepted

{Then, the} %so-called
azimuthal yield, $Y(\Delta \phi)$, is of particular interest, being defined by integration over a given
$\Delta y$ range {as} %:
\begin{equation}\label{eq:yield}
Y(\Delta \phi) = \frac{\int_{  y_\text{inf} \leq |\Delta y| \leq y_\text{sup} } S(\Delta y, \Delta \phi) dy}{\int_{y_\text{inf} \leq |\Delta y| \leq y_\text{sup}} B(\Delta y, \Delta \phi) dy},
\end{equation}
where {$y_{\rm inf(sup)%/sup
}$} defines the lower {(upper)}%/upper %MDPI: rewritten, fits better, makes it clearer. Please consider.   %Author: accepted
~integration limit for different rapidity intervals depending on the kinematic region of interest.
{\color{black} The $S(\Delta y,\Delta\phi)$ {\eqref{eq:same_evt}} %MDPI;added to clarify to the readers. Please consider; same for the B-function just below. %Author: accepted
used for the yield calculation are defined {as}
%follows, %MDPI: 1. No subequations in this journal.  2. the subscripts moved to the regular (Roman/normal) font as are the word abbreviations, otherwise mistakenly looking a product of the variables. Please consider.  %Author: accepted
%\begin{subequations}
\begin{equation}
%\label{eq:yield2}
%\begin{align}
S^{\rm SM}(\Delta y,\Delta\phi)=\frac{1}{N_\text{pairs}^{\rm SM}} \frac{d^{2}N^{\rm SM}_\text{same}}{d \Delta y d \Delta \phi}
\label{eq:yield2a}
%\\
\end{equation}
or
\begin{equation}
S^{\rm SM+HV}(\Delta y,\Delta\phi)=\frac{1}{N_\text{pairs}^{\rm SM}+N_\text{pairs}^{\rm HV}} \left(
\frac{d^{2}N^{SM}_\text{same}}{d
\Delta y d \Delta \phi}+
\frac{d^{2}N^{\rm HV}_\text{same}}{d \Delta y d \Delta \phi} \right) \label{eq:yield2b}
\end{equation}
%\end{align}
%\end{subequations}
{for} %MDPI: moved here, after Eqs., as beter suits and makes it clearer so. Please consider. %Author: accepted
the cases {when there} %MDPI; fits better. Please consider.
%in which we have
only the SM assumption {is put forward} or also the assumption of the HV existence, {respectively,}
with $B(\Delta y,\Delta\phi)$ {\eqref{eq:mixed_evt}}
being common for both as it is estimated from an enriched SM back-to-back sample, as defined
%in the text.
{in Section \ref{sec:results} below.
}

\section{Hidden Valley Phenomenology}
\label{sec:hv_pheno}
In most HV models \cite{Strassler:2006im}, the SM gauge group sector $G_\text{SM}$ is extended by (at least) a new gauge
group $G_\text{V}$ under which all SM particles are neutral. Hence, a new category of {HV} %MDPI: clarifies to te readers why "v"-prefix appears. Same below for $v$-quarks. %Author: accepted
%\textit
{$v$-particles}}
%%MDPI: We removed the full italics as not necessary as it  looks, mismatches while along with "v" as a product of variables. Meantime, the italics is reserved for the book titles, section names, volumes of journals etc. Please consider. Please confirm if the italics in full text are necessary; if not, please remove them. %Author: accepted
emerges charged under $G_\text{V}$, but neutral under $G_\text{SM}$. Moreover, {`communicators',} %MDPI; single-quotes enough.   %Author: accepted
charged under both $G_\text{SM}$ and $G_\text{V}$, are introduced to the theory allowing interactions between SM and HV particles. Communicators can be either intermediate 
{(considerably} %MDPI; or "considerably", "especially".  %Author: accepted "considerably"
%very %MDPI: No "very", "extremely", "hugely", "vastly" etc. recommended for scientific papers. Please consider replacing. Here and elsewhere below.
massive) vector $Z_v$ bosons, or
hidden partners of the SM quarks and leptons, generically denoted as $F_v$, assumed to be pair-produced. In this study, $F_v$ are
%taken as
{considered} %MDPI: fits better. Please consider.   %Author: accepted
fermions ({of the spin %=
\half}) while
{HV} $v$-quarks ($q_v$) %as
{considered} scalars.

Different mechanisms can {be considered to} %MDPI; missing wordin added. Please consider. %Author: accepted
connect the hidden and SM sectors through communicators, {for exmaple, }
%e.g.,
via the tree-level channel: $e^+ e^- \to Z_v \to q_v\Bar{q}_v \to$ hadrons. Alternatively, communicators
can be pair produced via SM $\gamma^{\ast}/Z$
coupling to a $F_v\bar{F}_v$ pair, yielding both visible and invisible cascades in the same event.
We have checked that the latter channel significantly dominates over the former within the range of energies under~study.

For some values of the HV parameter space, communicators {may} %can
{%\em
promptly} decay into a particle $f$ of the visible sector (its SM
partner) and a $q_v$ of the hidden sector according to the splitting: $F_v \to f q_v$. Note that the $q_v$ mass may {quite}
strongly influence the kinematics of the visible cascade (leading to SM particles) in the same event, thereby yielding
observational consequences. This mass remains currently unconstrained ranging from zero to close to the $F_v$
mass~\cite{Perez-Ramos:2021mvm}, so different values of {$F_v$} %it
{are %will be
assumed %in our analysis.
throughout this study}.

Indeed, we are especially interested in
the influence of the invisible HV sector
on the partonic cascade into final-state SM particles, leading to potentially observable effects. This
can be  understood, {for example,}
%e.g.,
as both visible and hidden cascades have to share the event's total energy, thereby
modifying their respective available phase spaces. For
the sake of
concreteness, we restrict {this} %our
study to a QCD-like HV scenario with $q_v$-quarks, $g_v$-gluons, and
an equivalent {`strong'} coupling constant $\alpha_v$.
For simplicity, $\alpha_v$ is assumed not
running with energy but fixed to $\alpha_v = 0.1$, as no significant differences {are} %were
found in
%our analysis
{this study} by varying {that $\alpha_v$} %this
value.

{\color{black} Present limits on masses and couplings in hidden sectors have been set by experimental searches at hadron colliders,
particularly the {LHC}~\cite{Lagouri:2022ier}.  %MDPI: please provide a justifying, clarifying ref/s (a review/s) would be higly  helpful the  readers can follow.
However, these bounds are primarily derived from searches targeting long-lived particles, events with
{relatively} large %MDPI: No "small", "big", "large", "fast", "slow", "near", "strong", "far", "drastic" etc.  with no comparison, estimate recommended for scientific papers. Please consider. Here and elsewhere below. %Author: accepted
missing transverse momentum, or relying on mass peak reconstruction. As such, %they
{bounds}
focus on specific signatures that probe only
certain realizations of the HV scenario. For example, current limits on {BSM  $Z'$ vector bosons
%beyond the Standard Model
exclude}
masses below 2–4 TeV. However, as {noted just  %discussed
above,}  these limits do not apply
%to our analysis.
{here.}

%In fact, %MDPI: Colloquial, highly not recommended for scientific papers. Please consider replacing or removal. Here and elsewhere below.  %Author: accepted
{Acually,}
a broad class of models falls under the {`umbrella'} of the HV framework, ranging from {%\em
unparticles}
\cite{strassler2008unparticlemodelsmassgaps} to soft bomb–like phenomena \cite{Knapen_2017}, dark sector showers \cite{Albouy_2022}
and dark matter candidates \cite{Beauchesne_2019}. Moreover, since the methodology proposed in this
%work
{paper}
notably differs from conventional search strategies, much of the parameter space remains largely unconstrained.

In summary, the values chosen throughout {the present study} %this paper
for the masses of the $v$-quarks, communicators, and the coupling constant $\alpha_v$ can be regarded as benchmark scenarios that illustrate the expected behavior of angular correlations and the feasibility of~detection.}

\section{Analysis at Detector Level}
\label{sec:results}

To assess the feasibility of HV signal detection
in $e^+e^-$ colliders, we {rely} %haverelied
throughout on the \Pythia\ Monte Carlo event generator~\cite{Sjostrand:2014zea} because of its {\color{black}
%well-} %MDPI: No "well/wide/etc-known/used/etc" recommended for scientific papers. Please consider. Here and elsewhere below.  %Author: accepted
{established}
reliability and the
{feature} %MDPI: or "feature".   %Author: accepted feature
%fact
that the HV production channel is already built in.

Fast detector simulations were performed using the \textsc{SGV} tool~\cite{berggren2012sgv} and the geometry and acceptance
{\color{black} according to the {extended} %large
version of the} model of the ILD concept for the International Linear Collider (ILC), as described in
Ref.~\cite{ILDConceptGroup:2020sfq}.
The simulated events {\color{black} were} provided in the same event model by the ILD concept group, and
the tools available in the {\textsc{ILCSoft} package}~\cite{Gaede:2006pj} %MDPI; please provide te hcorresponding ref. to help the readers to follow.  %Author: added, also in the .bib file
{%blue %MDPI: looks a remnant of the changes applied. Please confirm the no-need.   %Author: indeed, the blue is a remmant of previous versions. Not needed.
were} used for the event reconstruction and analysis. The ILD
reconstruction is based on {the %so-called  {%\em
particle flow %Particle Flow
} approach~\cite{Thomson:2009rp}, %MDPI: please add clarifying ref, otherwise teh eraders are lost here, one cannot refer to the approach, model etc. w/o correcponding ref. %Author: added
which {is considered a tool} %aims
to reconstruct
all individual particles produced
in the final state via pattern recognition algorithms. {Then, the} reconstructed candidates {\color{black} of single} particles are called
%Particle Flow Objects
{particle flow objects}
(PFOs). %MDPI: please add justifying ref. %Author: we don't think that it is needed, since we added the reference just above.

In {this} %our
study, {\color{black} which includes} detector effects, we set the centre-of-mass energy {(c.m.)} %MDPI: introduced to be used later on. Please consider. %Author: accepted
$\sqrt{s}$ of $e^+e^-$ collisions equal to
$250\;\mathrm{GeV}$ {%blue %MDPI: Please confirm is the remnant of teh changes to the text, no-need. %Author: confirm
corresponding to} the planned first commissioning of the collider as a Higgs boson factory.
As a further outlook, we {extend}  %ed %MDPI: as done here. Please consider here and elsewhere below.  %Author: accepted
our prospective study (this time only at particle level) up to $\sqrt{s}=500$~GeV and 1~TeV. The
case of longitudinally polarised beams
%will
{is planned to}
be addressed in %a
future %work
{investigations} %MDPI: plural fits better. Please consider.%Author: accepted
{as soon as} %MDPI: fits better in scientific papers. Please consider.  %Author: accepted`
%once
this option becomes available.

As already {notified in Section} {\ref{sec:hv_pheno},}
%commented in the previous section,
the HV signal proceeds via the process $\eee \to D_v\bar{D}_v\to \text{hadrons}$ (where $D_v$ denotes the lightest communicator,
being the hidden partner of the $d$-quark) as depicted in Figure~\ref{fig:diagrams}a. In {the}
%our
benchmark scenarios {considered here} we set
$\alpha_{v}=0.1$, $m_{D_v}=125~\gev$, with four different $v$-quark mass values: $m_{q_v}= 0.1, 10, 50$ and 100~\gev. We  also
consider $m_{D_v}=80$ and $100~\gev$ with $m_{q_v}= 40$ and 50~\gev  ~respectively. At higher {c.m. energies,}
%$\sqrt{s}$,
{\color{black} other
scenarios are also contemplated (see Section \ref{sec:higherenergies}).}

\vspace{-12pt}
\begin{figure}[H]
\subfloat[Hidden {V}alley\label{fig:hv}]{\includegraphics[width=4.cm]{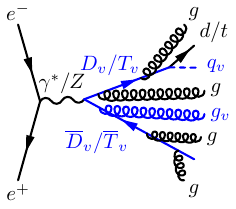} }
\subfloat[SM light quarks\label{fig:sm}]{\includegraphics[width=4.cm]{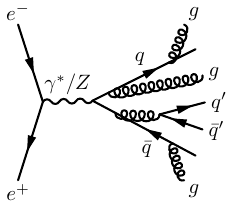}}
\caption{{Leading} %MDPI: We moved figure just after its first citation, please consider. %Author: accepted
diagrams of production in $e^+ e^-$ collisions for  (\textbf{a}) {HV %,
and}  (\textbf{b}) SM light quarks including bottom {quark}. %MDPI; missing wording, jargon. %Author: accepted
{See text for details.} %MDPI: otherwise to define all variables, quantities etc. as met first time.  %Author: accepted
\label{fig:diagrams}}
\end{figure}

In {the current study} %our analysis
at $\sqrt{s} = 250~\gev$, the $\eee \to q \bar{q}$ background arises from the inclusive production of all
%Standard Model
{SM}
quark
species except for the top quark, whose production is kinematically forbidden at this energy (see Figure~\ref{fig:diagrams}).

In a preceding investigation~\cite{Musumeci:2023rzf}, the {HV} %aforementioned
signal and background were examined at particle level.
However, {\color{black} no %Initial State Radiation
{initial state radiation}
(ISR) was included}, {\color{black} although} it plays a crucial role in the
{study} %analysis
as %we shall see. %MDPI: highly colloquial, not recommended for a scientific paper, but talks, lectures etc. Please consider. %Author: accepted
{shown below. }
Four-fermion production (dominated by $e^+e^- \rightarrow WW$) was {\color{black} neither} considered, although
{this process} %its
contribution {was} %has since been
found to be negligible. In the present {study,}  %work,
we extend {the investigation} %our study
to detector level, now taking into account the effect
of ISR. {\color{black} This inclusion} enables the production of the large {enough}
cross-section process $\eee \to \gamma Z$, where the $Z$
boson is produced on-shell and {\color{black} subsequently decays}. The cross-sections for both the HV and SM  processes %have been
{were}
estimated {\color{black}} using \Pythia\ {generator,} as {shown}  %reported
in Table~\ref{tab:cross_sections}.

%\vspace{-8pt}

\begin{table}[H]
\small
\caption{{Inclusive} %MDPI: We moved Table after its first citation, please consider. %Author: accepted
cross-sections for $\eee \to D_v\bar{D}_v$, $\eee \to q \bar{q}$ and $WW\to 4q$ processes at $\sqrt{s}=250~\gev$, with different
assignments for the $m_{D_v}$ and $m_{q_v}$ masses. Efficiencies of the {selection criteria,} % described in the main text,
{and event-averaged} %average
charged-track multiplicities {along with} %and
%their
$\rm RMS$ {of the multiplicities} %,
are {also} shown. %MDPI: Please check intended meaning has been retained. %Author: accepted
{See text for details.}
\label{tab:cross_sections}}
%\newcolumntype{C}{>{\centering\arraybackslash}X}
\begin{tabularx}{\textwidth}{cCCCCC} %{CCCCC}
\toprule
\multirow{2}{*}{\textbf{Process}} & $\bm{m_{D_v}}$ & $\bm{m_{q_v}}$ & $\bm{\sigma_{\textsc{\textbf{Pythia8}}}}$
& \textbf{Efficiency} & \multirow{2}{*}{$\bm{\langle N_\text{\textbf{ch}}\rangle}$} \\
%& \textbf{[\gev]} & \textbf{[\gev]} & \textbf{[pb]} & \textbf{[\%]} &  \\
& \textbf{{(}\gev{)}} %MDPI; we moved the squared brackets to the parenthses, as to be in the tables and clear enough as are units. Pleas consider. %Author: accepted
& \textbf{(\gev)} & \textbf{(pb)} & \textbf{(\%)} &  \\
\midrule
$e^{+}e^{-}\to D_v\bar{D}_v$ & 125 & \phantom{00}0.1 & $0.13$  & 36 & $ 12.4 \pm 3.7$ \\
& 125 & 10 & $0.12$ & 36 & $ 12.4 \pm 3.7$ \\
& 125 & 50 & $0.12$ & 42 & $ 11.4 \pm 3.5$\\
& 125 & 100\phantom{0} & $0.12$  & 42 & \phantom{1}$6.5 \pm 2.1$\\
& 100 & 50 & 1.29  & 42  &  $11.1\pm 3.4$ \\
& \phantom{1}80 & 40 & 1.54   & 36  &  $18.0\pm 4.9$ \\
\hline
$e^{+}e^{-}\to q\bar{q}$ with ISR & & & $48$ & $\lesssim$0.01 & \phantom{1}$9.9 \pm 3.4$\\
$WW\rightarrow 4q$ & & & $7.4$ & $\lesssim$0.001   & $-$ \\
\bottomrule
\end{tabularx}
\end{table}

In view of {quite} the
small cross-sections for HV production {as}
shown in Table~\ref{tab:cross_sections}, specific cuts {need to} %must
be applied to maximise
background suppression while retaining as much of the signal as possible. Displaced vertices resulting from the production and
decay of long-lived particles are not considered in this study, as {the} %we
focus {is} on prompt decays of $v$-particles and their impact on
the partonic cascade leading to final-state SM particles. Following a similar strategy to that in Ref.~\cite{Irles:2024ipg}, a set
of
selection cuts has been defined. The final selection-cut efficiency, reported in Table~\ref{tab:cross_sections}, demonstrates a
{quite a} drastic reduction of the SM background, while the HV signal remains largely {preserved.} %MDPI: we removed the next repeated paragraph, please confirm. %Author: accepted

%In view of the   small cross-sections for HV production from Table~\ref{tab:cross_sections}, specific cuts have to be
%{\color{black} applied} to maximise the background removal while {\color{black}retaining as much of the signal as possible}.
%On the other hand, displaced vertices  {\color{black} resulting from} the production and decay of long-lived particles are not
%considered  {\color{black} in this study} as we focus on prompt decays of $v$-particles and their   {\color{black} impact} on the
%partonic cascade  {\color{black} leading to} final-state SM particles. Following a similar strategy  {\color{black} to that}
%in~\cite{Irles:2024ipg}, a set of selection cuts {\color{black} has} been defined. The final selection-cut efficiency, reported in
%Table~\ref{tab:cross_sections}, shows  a drastic reduction of the SM background while the HV signal  {\color{black} remains largely
%preserved.}

The {following} selection cuts are
applied  in this {study}: %analysis are the following. %MDPI: we moved to the items, fits better and helps in clarifying to teh readers. Please
consider.  %Author: accepted
\begin{itemize}
\item
{$\rm S1$:} %MDPI: we moved to the regular (Roman/normal) font as those are words, not variables, quantities. Please consider.  %Author: accepted
constraints %have been
{are}
set on the number of displaced vertices per jet ({to equal to}  %=
0) and the number of neutral and charged PFOs
({below} %$\leq$
22
and %$\leq$
{below} 15, respectively);
\item
{$\rm S2$:} cuts are applied on the reconstructed ISR photon candidates angle \mbox{($|\cos{\theta_{\gamma_{\text{ISR}}}}|<0.5$}) and
energy ($E_{\gamma_\text{ISR}}<$ 40 GeV);
\item
{$\rm S3$:} constraints on the di-jet invariant mass ($m_{jj}<130$~GeV) and on the energy of the most energetic jet ($E<80$~GeV);
\item
{$\rm B1$:}%,
~a thrust
{value
%larger than %MDPI: extra wording, the inequality says it all. Please consider.  %Author: accepted
$%(
T>0.95%)
$}
%was
{is} imposed.
\end{itemize}

The {combined
%$S1+S2+S3 $
S1, S2 and S3}
cuts are applied to the evaluation of $S(\Delta y, \Delta \phi)$ {\eqref{eq:same_evt}},
%while
{and}
for the evaluation of $B(\Delta y, \Delta
\phi)$ {\eqref{eq:mixed_evt}},
{a combination of S1, S2} %$\rm S1+S2$
{along with} %and
{$\rm B1$} are applied. The selection efficiencies for different combinations of these cuts applied
to different processes are {given} %displayed
in Table~\ref{tab:cross_cutflow}.

\begin{table}[H]

\caption{{Breakdown} %MDPI: We moved Table just after its first citation, please consider. %Author: accepted
of accumulated efficiencies of the selection{ criteria.} % described in the main text.
For simplicity, {the efficiencies} only {for} one HV model is {given,} %shown,
since results are
%very
{quite}
similar for {others %the rest:
(}$e^{+}e^{-}\to D_v\bar{D}_v$ with $m_{D_v}=125~\gev$ and $m_{q_v}=100~\gev${)}.
The {results for the } two SM processes {(see}
%as  described in the text and  in
Table~\ref{tab:cross_sections}{)} are also {given.} %shown.
%The definition of the cuts, $S1+S2+S3$ and $B1$ are defined in the text. In this table we also
{Here, we} %differentiate
{distinguish} %MDPI; fits better, the former may be mismatched with the mathematical action. Please consider. %Author: accepted
between the $q\bar{q}$-RR (dominated by radiative return events) and the $q\bar{q}$-{c}%C%
ont. in the continuum (i.e., {with} $m^{\rm inv}_{q\bar{q}}\approx%\sim %MDPI; not "of an order" of but "about". Please consider.
\sqrt{s}$ at partonic level and before parton shower. %Author: accepted
{See text for further details.}
\label{tab:cross_cutflow}}\vspace{-6pt}
%\newcolumntype{C}{>{\centering\arraybackslash}X}
\begin{adjustwidth}{-\extralength/4}{0cm}
\begin{tabularx}{\fulllength}{lCCCC}
\toprule
& \multicolumn{4}{c}{\textbf{Efficiency {(}%[
{\%}{)}%]
}}  \\
\multicolumn{1}{c}{\textbf{Cuts}}
& \textbf{HV} & \boldmath{$q\bar{q}$}\textbf{-RR} & \boldmath{$4q$} & \boldmath{$q\bar{q}$}\textbf{-{c}%C%
ont.}\\
\midrule
S1{,}$~ %~ ~ ~ ~ ~ ~ ~ ~ ~ ~ ~ ~%=
$ %M
{i.e., multiplicity cut} & 89 & 30 & 64 & 40\\
%Multiplicity & & & & \\
S1 {and} %+
S2,$~ %~ ~ ~ ~ ~ ~%=
$ {i.e.,} S1 {and} %+
ISR photon {cuts} & 42 & 3.8 & 12 & 8.8 \\
%ISR photon  &  &  &  & \\
{S1, %+
S2 %+
and S3,}$~%=
$ {i.e.,} S1{,} %+
S2{, %+ %E
energy} and invariant mass  {cuts}& 42 & $\lesssim$0.01 & $\lesssim$0.001 & $-$ \\
%Energy and invariant mass &  &  &  & \\
{S1, %+
S2 and %+
B1,}$~%=
$ {i.e., S1, %+
S2 and} %+
thrust  {cuts}
& $-$ & $-$ & $-$ & 5.6\\
%Thrust & & & \\
\bottomrule
\end{tabularx}
\end{adjustwidth}
\end{table}

Figure~\ref{fig:correlations} shows three-dimensional plots of the two-particle correlation function $C^{(2)}(\Delta y, \Delta
\phi)$
{\eqref{eq:corr_function}}
%separately %MDPI: extra wording, removed, clear enough. %Author: accepted
for the %HV and SM
{SM  (Figure~\ref{fig:correlations}a) and HV (\mbox{Figure~\ref{fig:correlations}b})
scenarios,   before and after cuts.}
%before (Figure~\ref{fig:correlations}a) and after (\mbox{Figure~\ref{fig:correlations}b})
%cuts. %MDPI: wrong description, corrected so syas what is indeed shown in Fig 2. %Author: accepted
%As
{For}
a reference {sample %example %,
in} the HV {scenario,} %case
we set $m_{q_v}=100 \;\mathrm{GeV}$, $m_{D_v}=125~\gev$ and $\alpha_v=0.1$ alongside the  decay $D_v \to d + q_v$ initiating the
partonic (both visible and invisible) shower.
As expected, a near-side peak shows up at ($\Delta y \simeq 0 $, $\Delta \phi \simeq 0$), receiving contributions mainly from track
pairs within the same jet. On the other hand, an away-side correlation ridge around $\Delta \phi \simeq \pi$, and extending over
a {relatively} large rapidity range, results from back-to-back momentum balance,
%in principle %MDPI: Colloquial, highly not recomemnded for scientific papers. Please consider. Here and elsewhere below.    %Author: accepted
{actually,}
%un
{not} related to NP effects.
After cuts, a near-side ridge (with two pronounced and symmetric bumps) shows {itself} %up
for $1.6 < |\Delta y| < 3$ at $\Delta \phi \simeq 0$ for the SM {scenario,} %case,
similar to {that in} the pure HV scenario.
This structure arises from the ISR effect, since the effective {c.m.} %center-of-mass
energy approaches the $Z$ mass, and thus {the} resonant production
%greatly %MDPI: No "great/ly/er", "huge/ly", "extreme/ly" etc. recommended for scientific papers. Please consider replacing. Here and elsewhere below. %Author: accepted
{highly}
enhances the production cross-section. It becomes of paramount importance to take this effect into account in
%our analysis,
{this study} mimicking possible HV signatures in angular correlations.

\vspace{-8pt}

\begin{figure}[H]
%\centering
%\begin{adjustwidth}{-\extralength}{0cm} %MDPI: added to widen the figure. Same for other figures.  %Author: accepted
\subfloat[\centering]{
\includegraphics[width=4%4
.0cm]{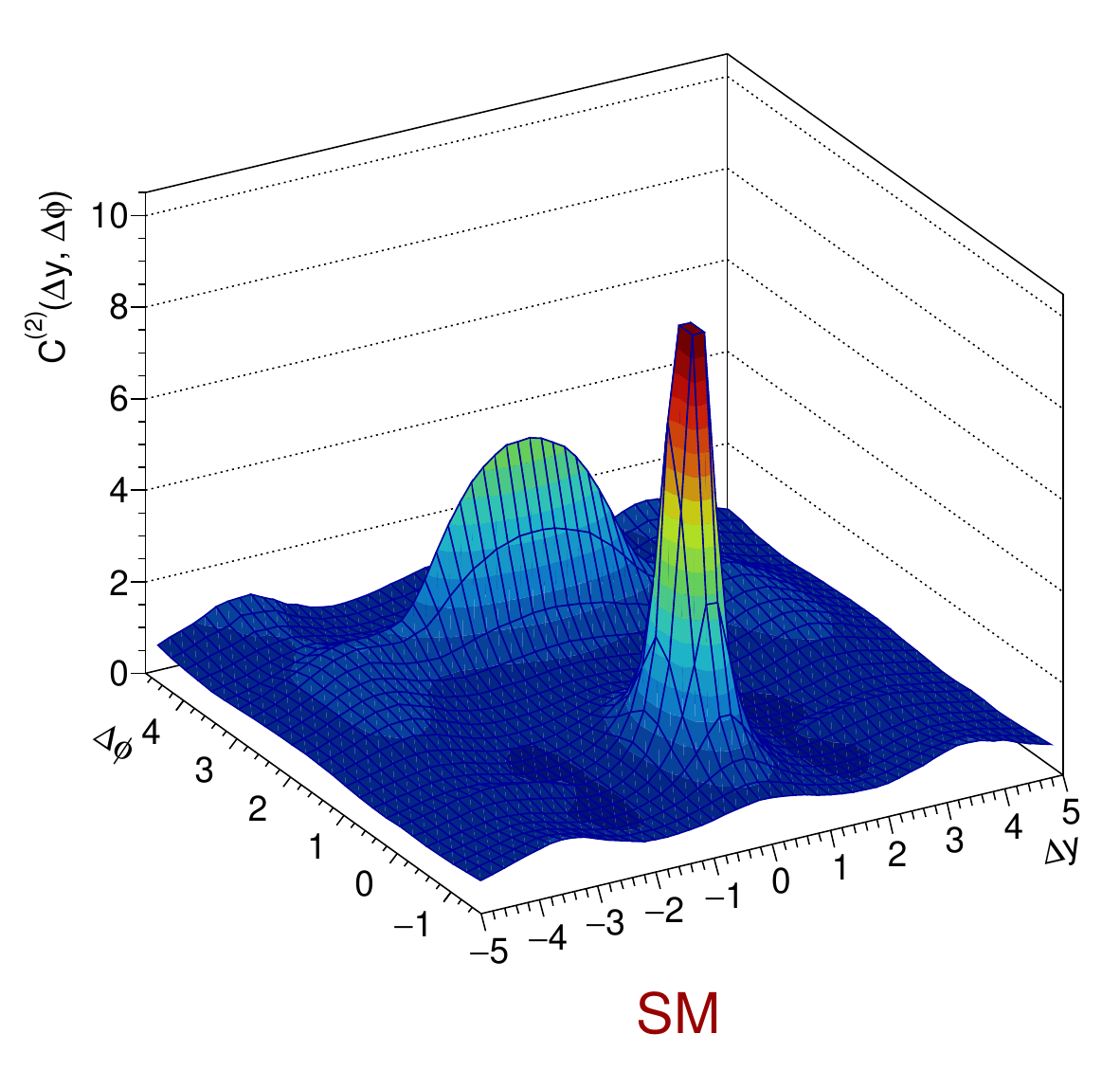}
\includegraphics[width=4.%4.5
0cm]{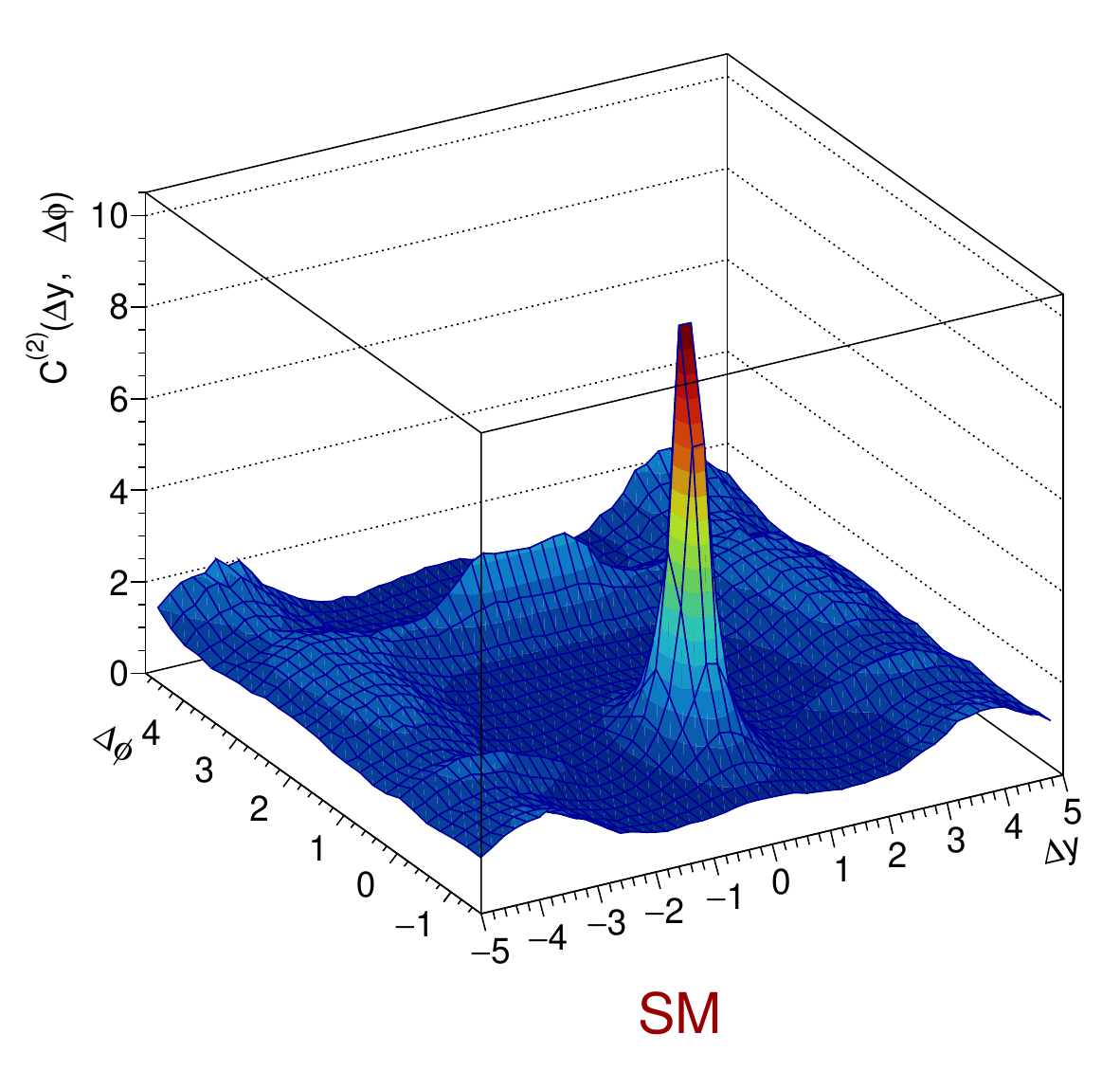}
\includegraphics[width=4.%4.3
0cm]{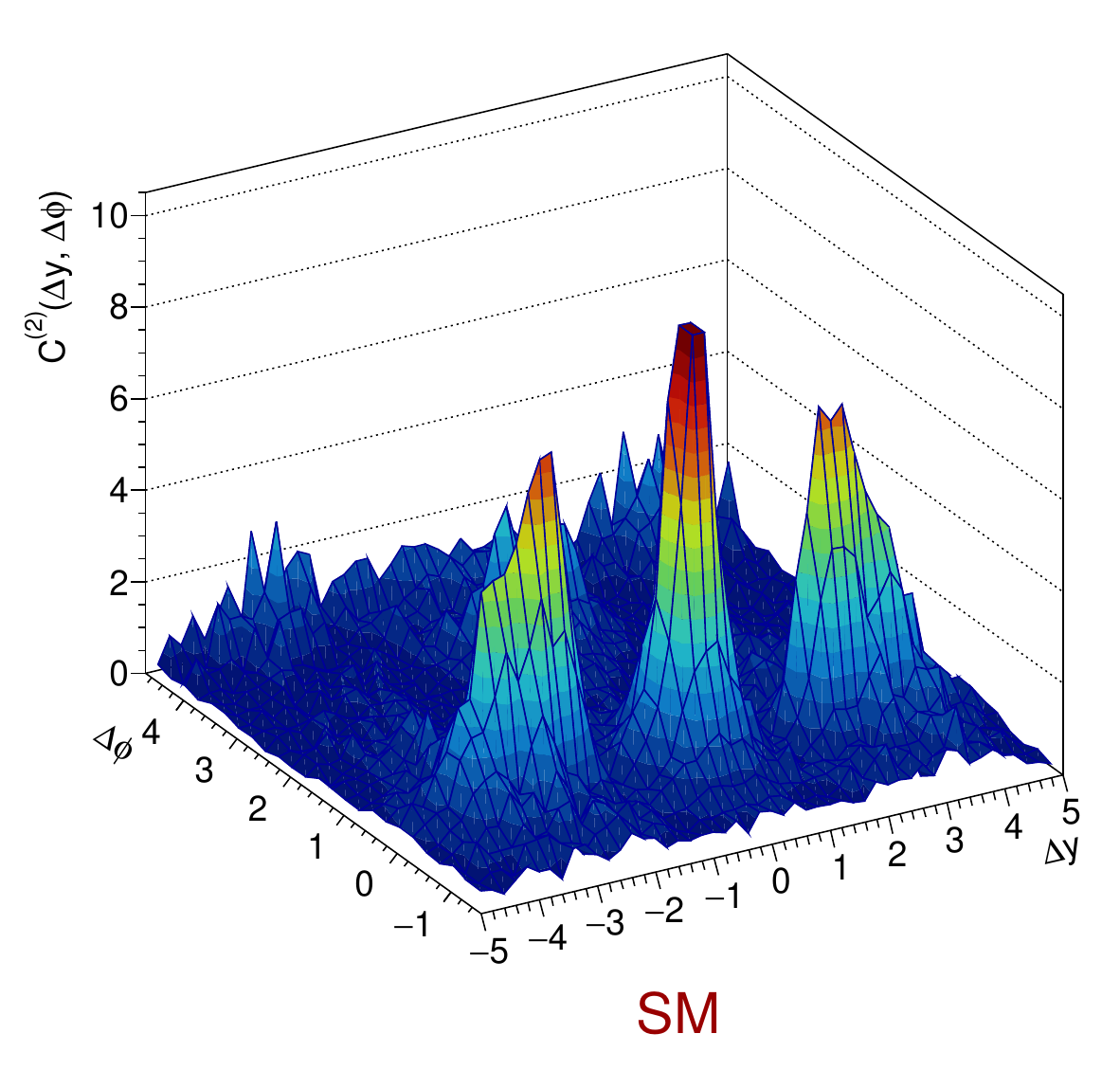}
}
%       \caption{\emph{{Cont}.}%MDPI: moved to be not shifted to the left.
%}
%\label{fig:correlations}
%\end{adjustwidth}
% \caption{\emph{{Cont}.}%MDPI: size of subfigures incresed to be better visible, same for other figures. Please consider. %Author: we are not sure if the "Cont." is a mistake or not. It seems strange to have twice the Figure 2 caption and the first of them being only with the word "Cont.". Is this intended?
%}
%\end{figure}

\vspace{-12pt}

%\begin{figure}[H]\ContinuedFloat

\hspace{60pt}\subfloat[\centering]{\includegraphics[width=4.%5
00cm]{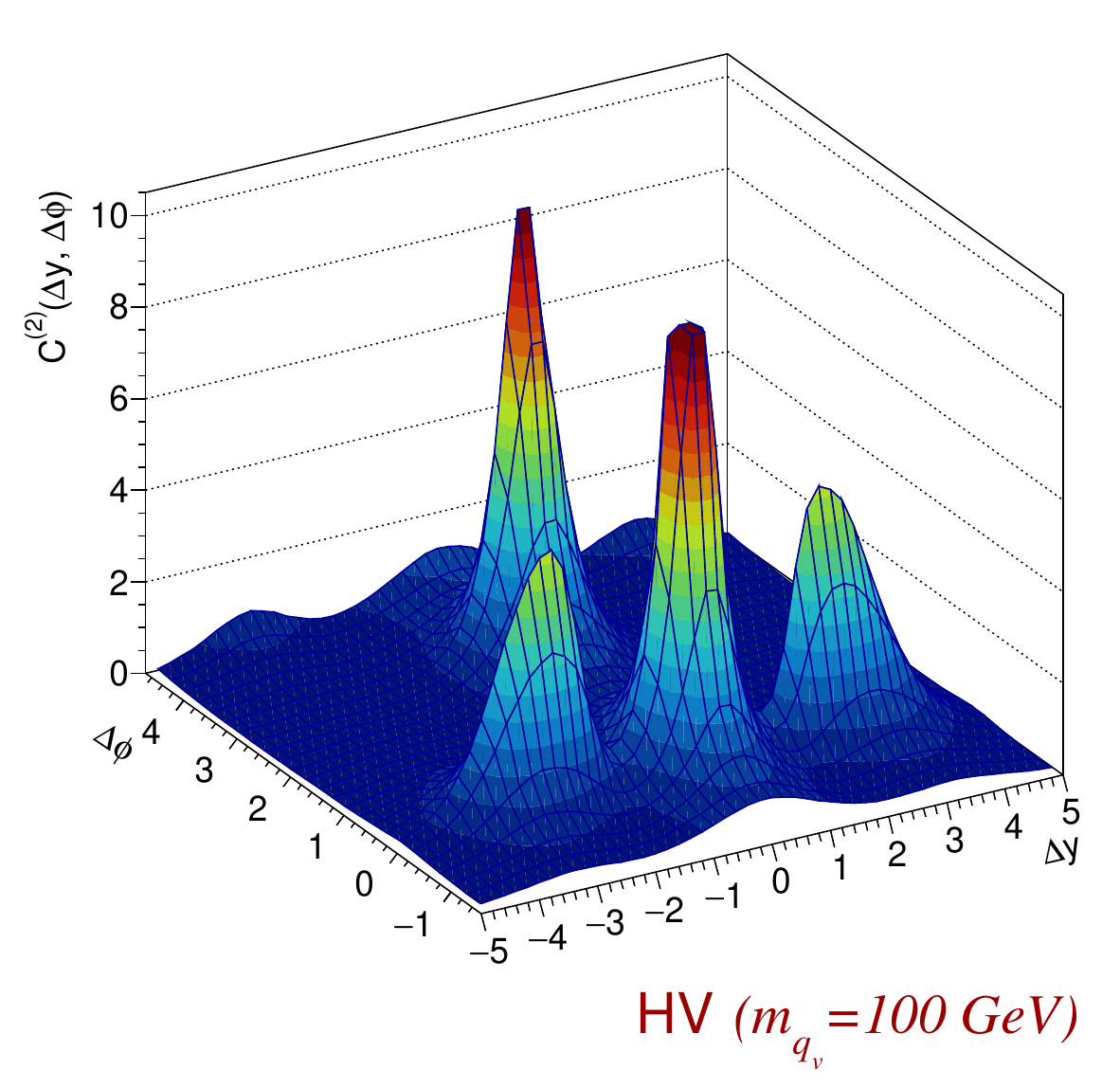}
\includegraphics[width=4.%5
00cm]{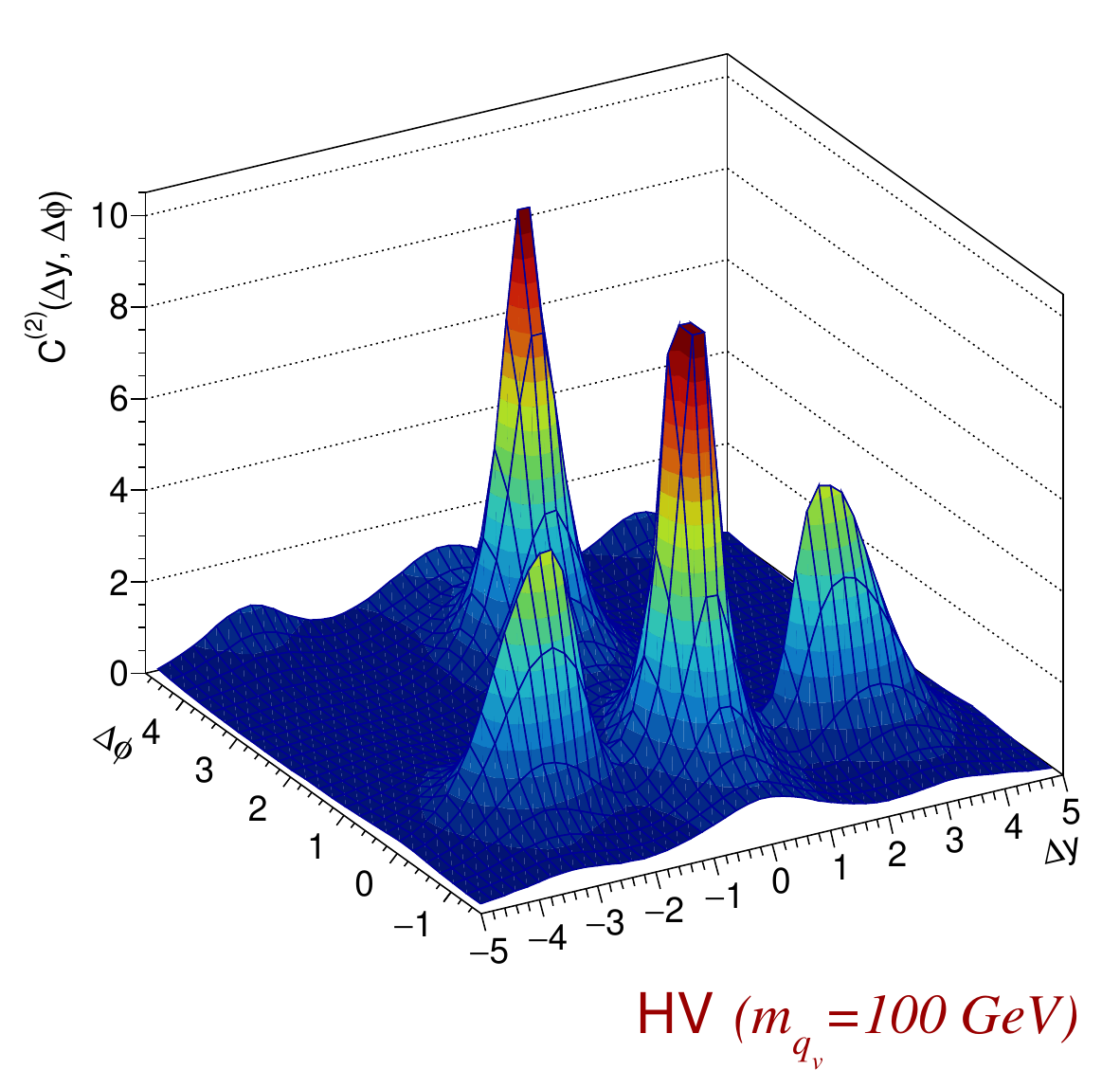}
}\vspace{-3pt}
% \end{adjustwidth}
\caption{{Three-dimensional} %MSPI: fits better, moreover not defined. Please consider. %Author: accepted
%3D
plots of the two-particle angular correlation function, $C^{(2)}(\Delta y, \Delta \phi)$ {\eqref{eq:corr_function}},
{built} at detector level {using \Pythia\ generator},
%separately %MDPI: extra wording, clear enough. Please consider.  %Author: accepted
{for}  %MDPI: reordered as the figure shows it. %Author: accepted
{({\bf a})} the SM background and  {({\bf b})} the pure HV signal
in \eee collisions at $\sqrt{s}=250~\gev$, {with} %MDPI: We moved the subfigure explanations into the figure caption and rewritten those what makes those clearer and shorter. Please consider.  %Author: accepted
{(\textbf{left})} no,  {(\textbf{a}, \textbf{middle})}  %$S1+S2$,
{S1 plus S2} and {(\textbf{right})} %$S1+S2+S3$
{combined S1, S2 and S3} cuts applied
to the $S(\Delta y,\Delta\phi)$ {\eqref{eq:same_evt}} reconstruction.
For the pure HV modelling:  $m_{q_v}=100~\gev$  light quarks include the bottom flavour; no  %$S1+S2$
{S1 plus S2} case is shown as no differences with
other cases are distinguishable. {See text for details.}}
\label{fig:correlations}
\end{figure}

In order to examine in more detail the possibility of discriminating the HV signal from the pure SM background, we depict in
Figure~\ref{fig:yield} the yield $Y(\Delta \phi)$ (defined in Equation~(\ref{eq:yield})) for both $0<|\Delta y|\leq 1.6$ and $1.6<|\Delta y|<5$ ranges:
before (Figure~\ref{fig:yield}a) and after (Figure~\ref{fig:yield}b) cuts. Note that the HV signal for various masses
$m_{D_v}$,$m_{q_v}$  is considered together with the SM background, while a standalone
%analysis
{findings} of the SM background %is %Author: not findongs, findings
{are} also presented.
A {visible} %clear
difference becomes apparent for $0<|\Delta y|\leq 1.6$: a sizeable peak at $\Delta \phi \approx %\sim
\pi$ characterises the HV scenario, unlike the pure SM case. This remarkable discrepancy of shapes
%would
{may}
potentially serve as a valuable signature of a hidden sector,
complementary to more conventional BSM searches, as claimed in this study.

\begin{figure}[H]
\vspace{-12pt}\centering
\begin{adjustwidth}{-\extralength/4}{0cm}
\centering
\setcounter{subfigure}{0}
\subfloat[\centering]{
\includegraphics[width=8%5
.00cm]{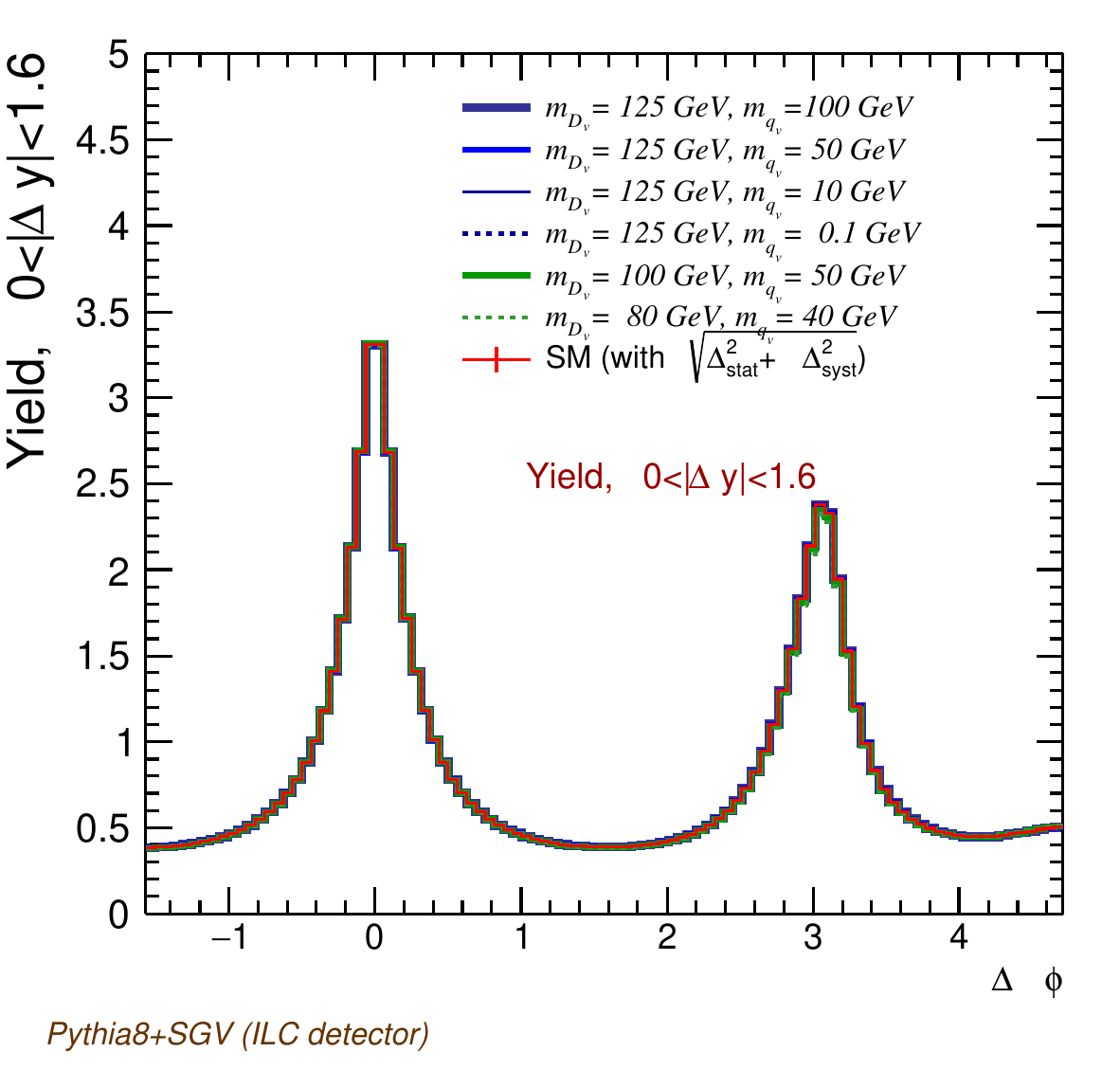}
\includegraphics[width=8%5
.00cm]{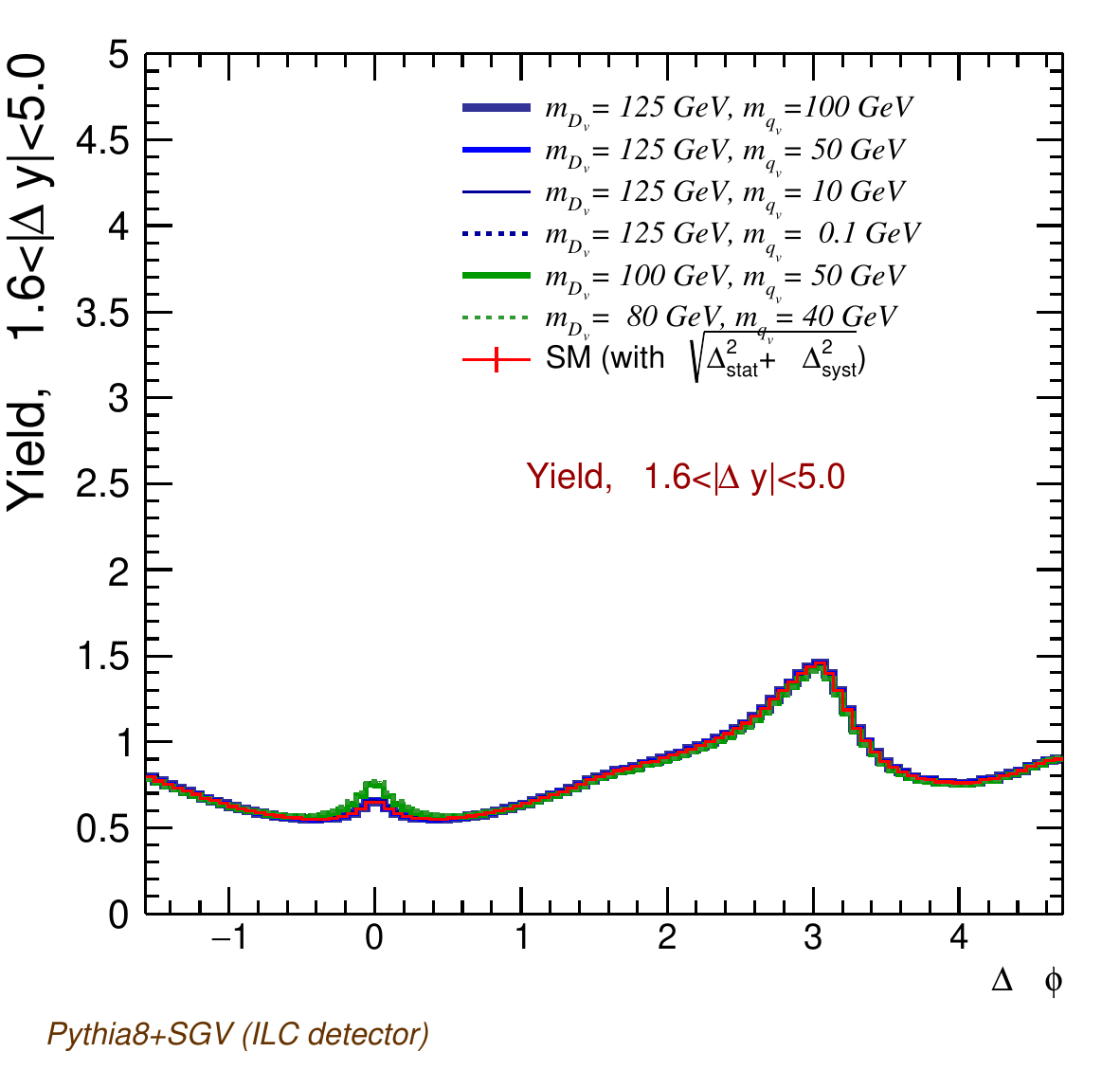}}\\
\vspace{-12pt}
\subfloat[\centering]{
\includegraphics[width=8%5
.00cm]{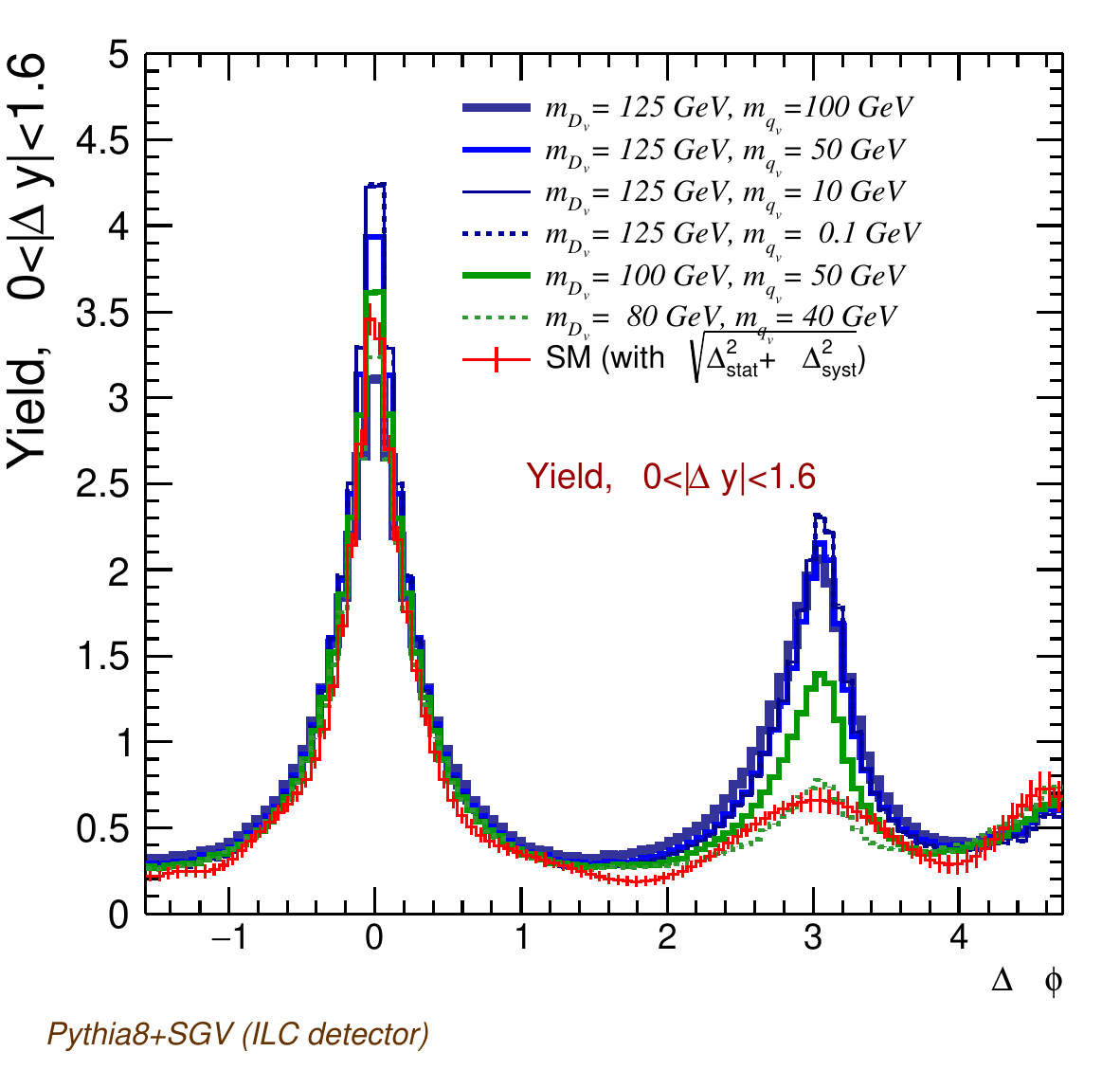}
\includegraphics[width=8%5
.00cm]{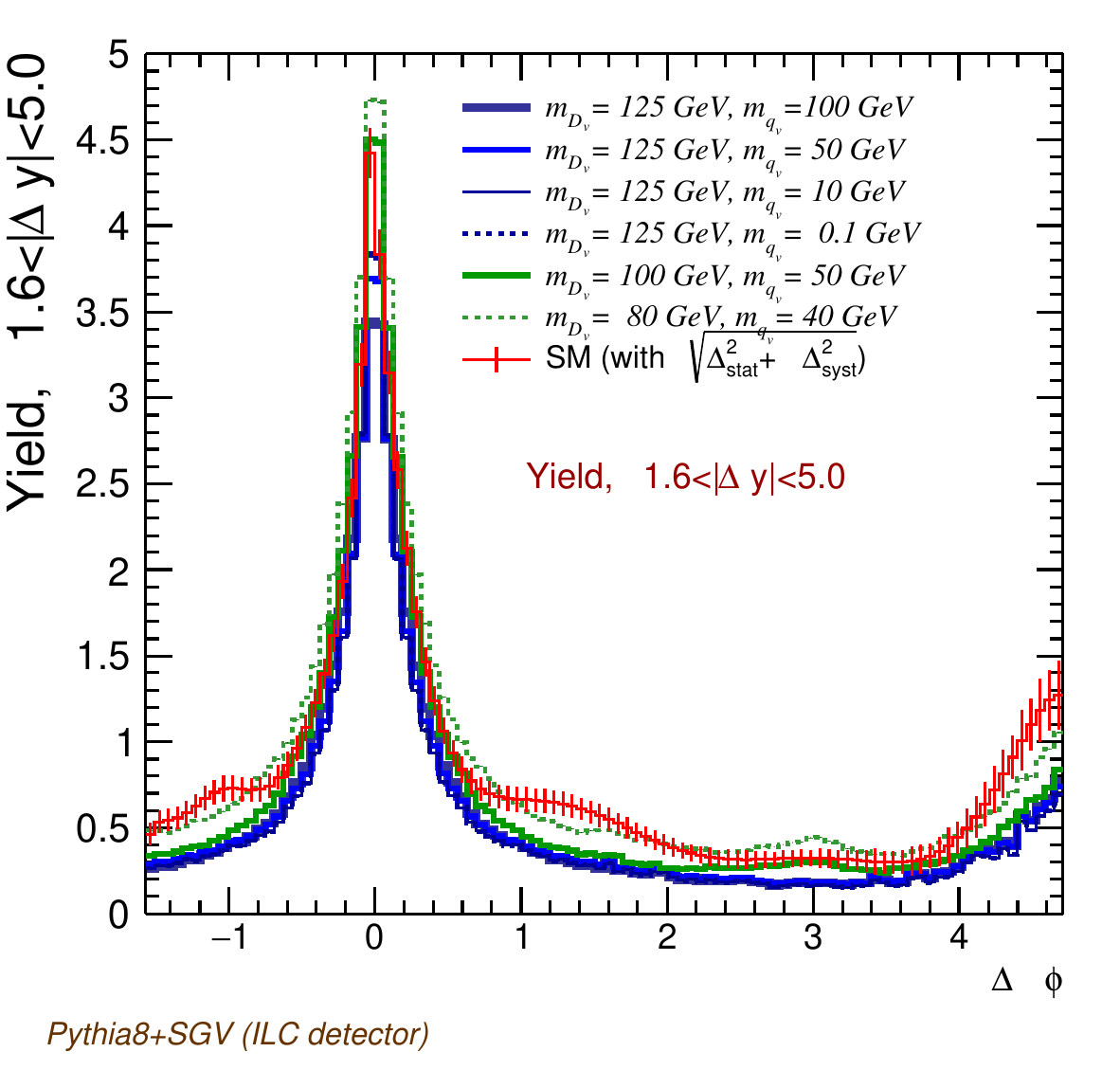}}
\end{adjustwidth}
\caption{\label{fig:yield} {Yield} %MDPI: the size of subfigures increased to be better visible, then shifted. Please consider. %Author: accepted
$Y(\Delta \phi)$ {\eqref{eq:yield}}{, built} at detector level %MDPI: We moved the subfigure explanations into the figure caption. Please confirm. %Author: accepted
{using \Pythia\ generator, }
{for} %both  %MDPI: Not necessary, clear enough as indicated. %Author: accepted
HV signal in association with SM {(\eqref{eq:yield2b}}
{(for different $m_{D_v}, m_{q_v}$ mass pairs, as indicated)}
and for pure SM background {\eqref{eq:yield2a}} (\mbox{red line}) for the $0<|\Delta y|<1.6$  {(\textbf{left})} and
$1.6<|\Delta y|<5$ {(\textbf{right})} intervals, %respectively %MDPI: No correspondence, so removed. Please consider.  %Author: accepted
%The $S(\Delta y,\Delta\phi)$ used for the Yield calculation are defined in Equations (\ref{eq:yield}),
%(\ref{eq:yield2a}) and (\ref{eq:yield2a}).
{without}
{(\textbf{a})} and with {(\textbf{b})} selection cuts applied.
Notice the different {between} shapes of the HV %+
{plus} SM signal and {pure} SM background resulting after applying the selection cuts, {so} providing a valuable
signature of NP using angular correlations.
The {pure-}SM curve also shows the error bars corresponding to the luminosity scenario {statistical ($\Delta_{\rm stat}$)}
and current state of the art %of
systematic {($\Delta_{\rm syst}$)}
uncertainties as discussed in Section \ref{sec:sens}.
}
\end{figure}

\section{Prospects on the Experimental Sensitivity}
\label{sec:sens}

Let us stress that {a %an  analysis to  %MDPI: not necesary, extra wording. Please consider. %Author: accepted
search} for BSM physics in high-energy collisions relying on angular correlations (as proposed in
this paper) offers several advantages with respect to more conventional methods, {for example}
%e.g.,
based on an excess of events in cross
sections or invariant mass peaks. In particular, yield {distributions %as defined in Equation %MDPI: not necessary as the characteristic wording is given. %Author: accepted
(\ref{eq:yield})} benefit from an almost
total
cancellation of reconstruction efficiencies and detector acceptances, luminosity and cross-section dependence, {and so on.} %etc.
However,
modelling uncertainties could be a limiting factor for these observables. In order to better understand this issue, we performed a
study assuming different scenarios for the foreseen statistic and systematic uncertainties.

To estimate the statistical uncertainties, we {assumed} a collected luminosity of 100~\ifb, which
%roughly %MDPI: colloquial, highly not recommended for scientific papers. Please consider. here and elsewhere below.  %Author: accepted
{approximately}
corresponds to one year of data taking of ILC in the H20-staged scenario~\cite{Bambade:2019fyw}.
For the systematic uncertainties, we identified two potential main sources: the detector performance modelling and the fragmentation
(pertaining to the final hadronisation of the partonic cascade) uncertainties. For both of {these scenarios,} %them,
only educated guesses based on
current knowledge can be made. For instance, the detector performance modelling uncertainty is evaluated by a bin-by-bin comparison
of the estimated yield distributions at particle and detector levels. The absolute value of the difference is taken as uncertainty.
{\color{black} The estimated size of this uncertainty is of the order of 5--10\% in each bin. This comparison should account for the
kinematic resolution (including angular) on the track reconstruction, as well as acceptance effects. All detectors proposals for
future Higgs %F
{factories} %MDPI; no capitalizing necessary. Please conider. Here and elsewhere below. %Author: accepted
foresee high-performing and low mass tracker systems, with near-$100\%$ tracking efficiency and
{transverse} momentum
resolution {of} $\sigma(1/p_{T})=2\times10^{-5}$~GeV$^{-1}\bigoplus 1\times10^{-3} %\text
{p}_{T}^{-1} %MDPI: p is the variable, so in the math mode. %Author: accepted
\sin^{-1/2}{\theta}$ or
better~\cite{ILDConceptGroup:2020sfq}. For those reasons, %we
{to
stress is} that this approach highly overestimates the experimental
uncertainties therefore, it represents a worst-case scenario.} For the fragmentation uncertainty, we {followed} the same recipe as for
the detector-effects modelling, but this time we {compared} the predictions of the yield at the particle level calculated using two
different fragmentation models implemented in \Pythia\ and \Herwig\ {generators}~\cite{Bahr:2008pv,Bellm:2015jjp}. %, respectively. %MDPI: nothing to correspind. Please consider.  %Author: accepted
{\color{black} In this case, the  estimations lead to a subdominant contribution of
the uncertainty of about a third of the estimated detector modelling uncertainties.}  The total uncertainty, calculated also bin by
bin, is composed of the addition in quadrature of the statistical uncertainty for the  expected number of events, and the systematic
uncertainties of {$Y_{\rm SM}$} since only \Pythia\ {generator}
includes Hidden Valley production. The  {simualtion results due to} outcome of these
assumptions {are} %is
summarised in Figure~\ref{fig:sensitivity}, with $\sigma_{Y}$ representing the estimated uncertainty as explained {just}~above.

%Once
{As soon as}
all the %above
points {discussed just above}
are taken into account, it is straightforward to calculate the sensitivity of the observable by comparing
the HV and SM predictions for $Y$-{yields} with the expected uncertainties. Already at 100~\ifb, the sensitivity is {found to be} %MDPI: as thsi seems be a result of the studies. Please consider. Otherwise looks like facts known from beyond the studues done here. Same just below. %Author: accepted
mostly limited by {the}
systematic uncertainties as the detector performance systematics is {found to dominate}
%dominant
over all the other {uncertainties.} %s.
However, the discovery power
remains almost intact, especially near the sizeable $Y$ peak at {$\Delta \phi \approx %\sim
\pi$} with $0<|\Delta y|\leq 1.6$. As an exercise,
we {compared} the estimated sensitivity with a much more optimistic scenario in which we {improved the} %our knowledge of
systematic
uncertainties by one order of magnitude with respect to the current estimates
assuming a significant improvement in detector-performance-related uncertainties. Dedicated studies on the fragmentation modelling
of HV processes {is to be %would
also} %be
required for further understanding. Assuming this order of magnitude improvement, the sensitivity %would  be drastically enhanced,
{considered to quite drastically increase}
and the discovery power {to reach} %would be reached
for most of the phase space available. %Of course, we
{Let us}
emphasise that this is only an educated guess that {to} %should
serve as yet another motivation to pursue the best possible design of
future collider detectors and progress on Monte Carlo tools development to minimise \mbox{modelling uncertainties.}

\begin{figure}[H]
\begin{adjustwidth}{-\extralength/4}{0cm}
\centering
\setcounter{subfigure}{0}
\subfloat[\centering]{
\includegraphics[width=9%6
.00cm]{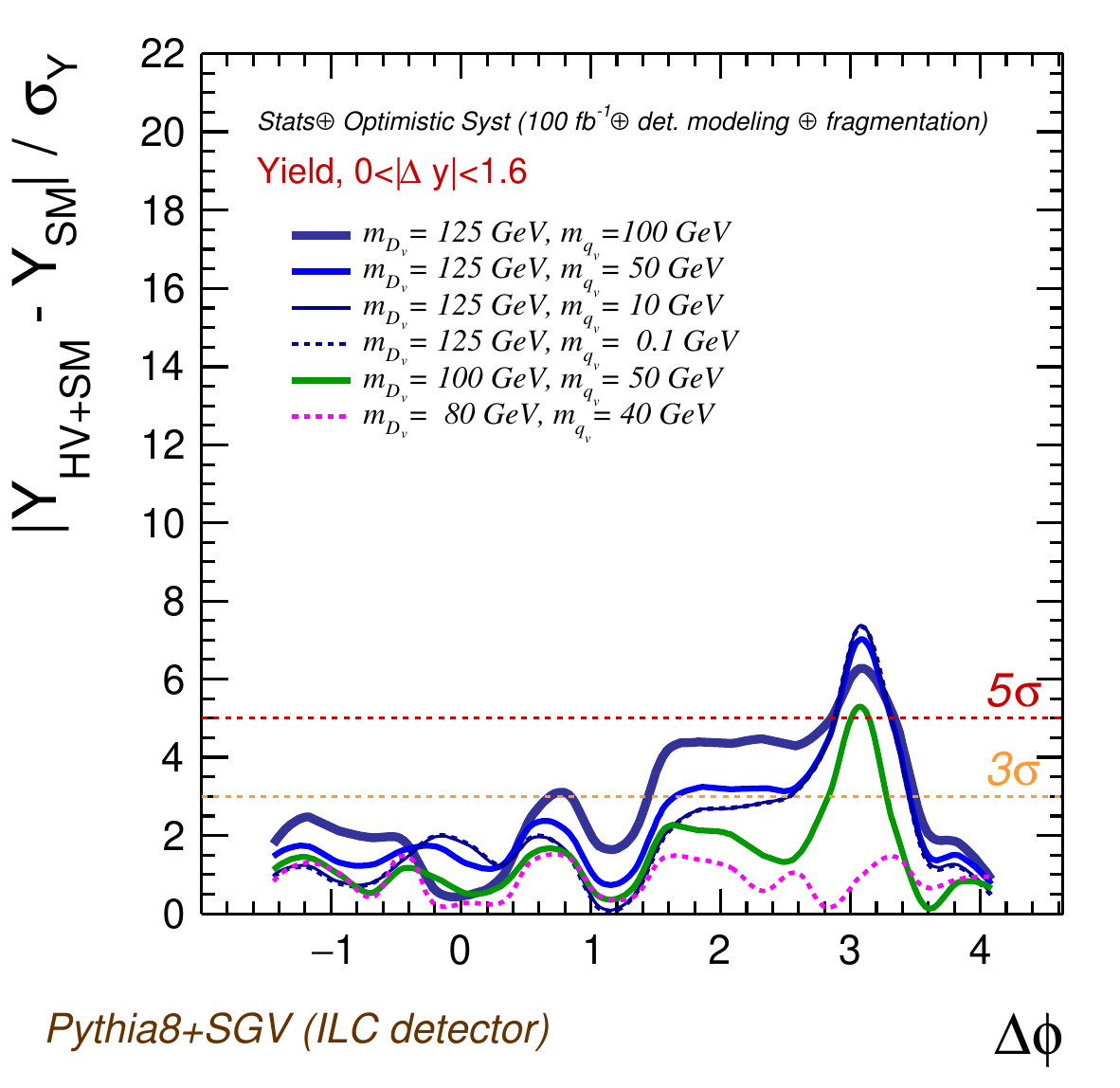}
\includegraphics[width=9%6
.00cm]{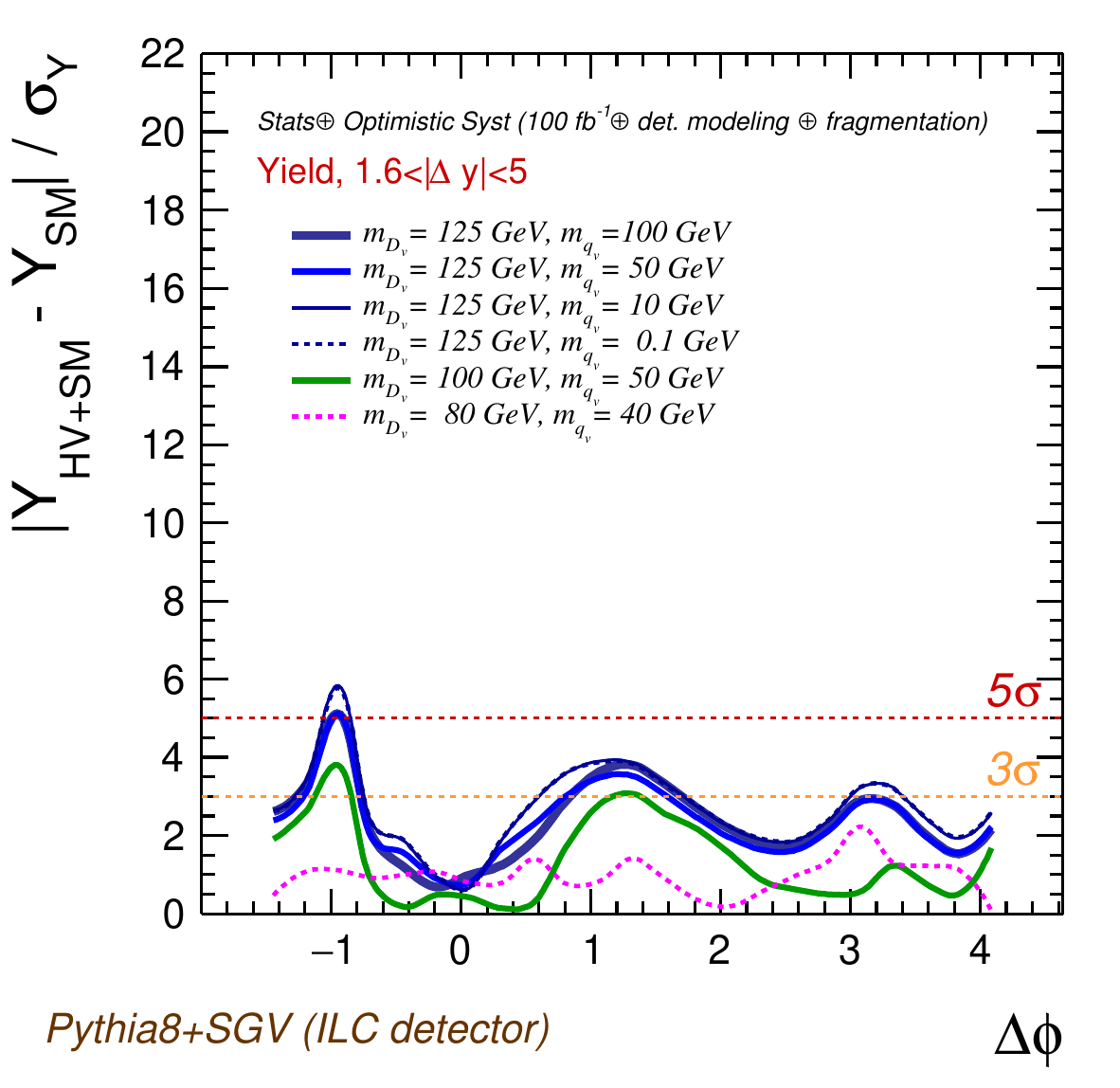}
}
\hfill
\subfloat[\centering]{
\includegraphics[width=9%6
.00cm]{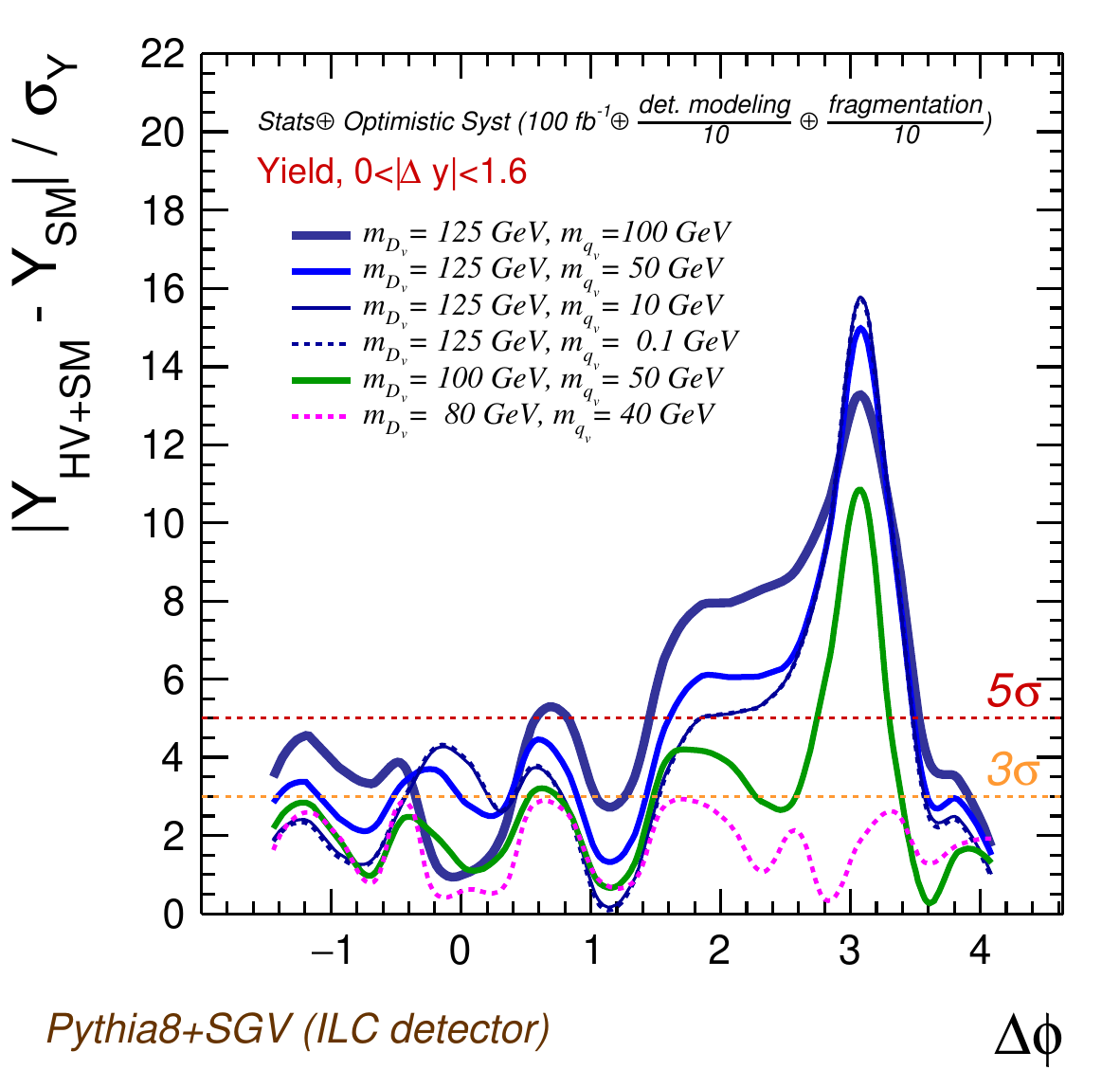}
\includegraphics[width=9%6
.00cm]{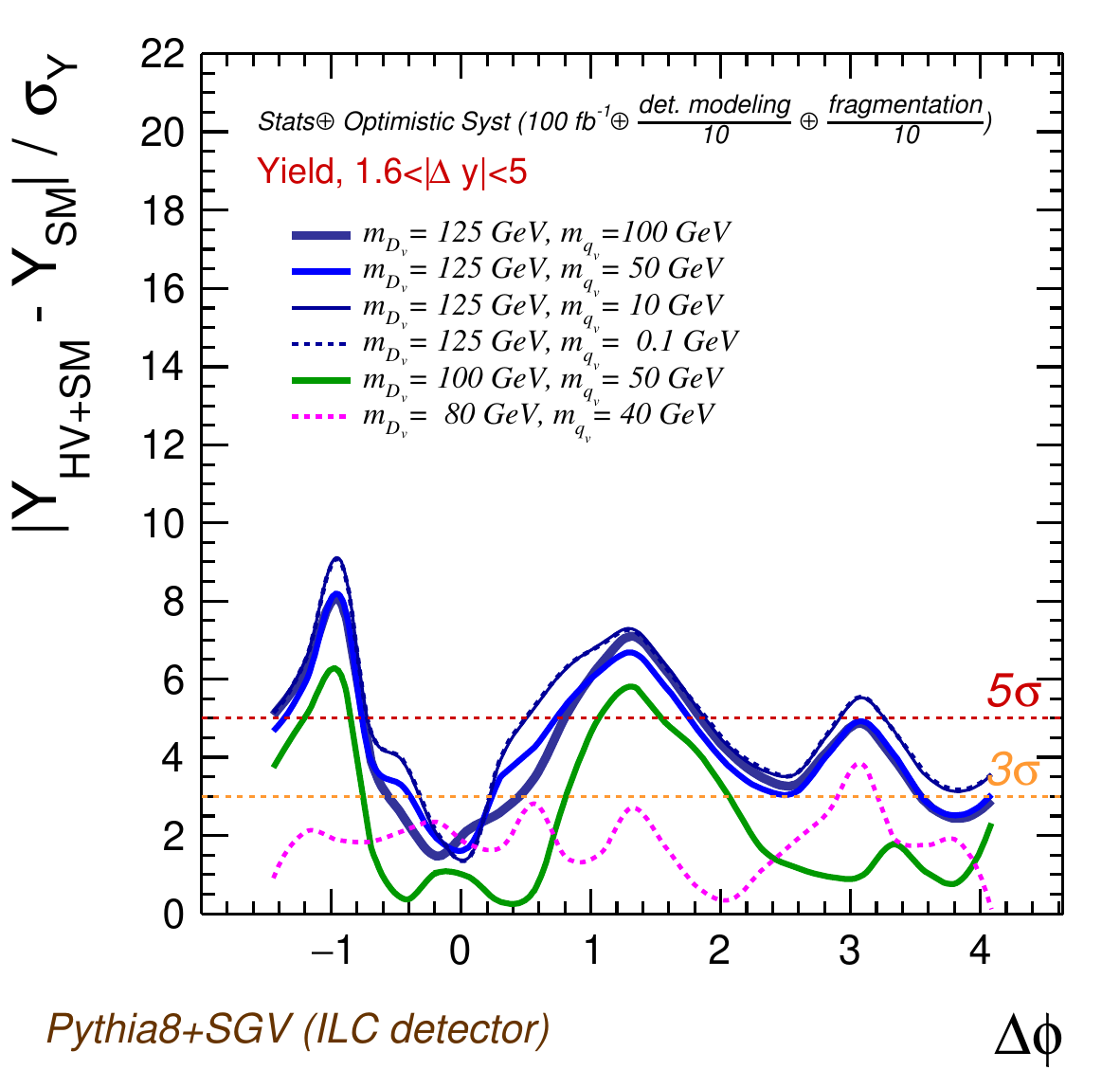}
}
\end{adjustwidth}
\caption{\label{fig:sensitivity} Expected experimental sensitivity for {HV} %Hidden Valley
models {compared to} %MDPI: thsi fits better as in accordance what is s hown as a difference. Please consider. %Author: accepted
%along with
the {pure} SM background after collecting
100~\ifb of integrated luminosity, for yield measurements in the range of \mbox{$0<|\Delta y|<1.6$}  {(\textbf{left})} and $1.6<|\Delta y|<5$
{(\textbf{right})}
{with} %MDPI: We moved the subfigure explanations into the figure caption. Please confirm. %Author: accepted
systematic modelling uncertainties obtained (\textbf{a}) from current state of the art and (\textbf{b})
assuming an improvement of one
order magnitude on the modelling \mbox{uncertainties evaluation}.
{The horizontal dotted lines show the exclusion at the level of 5 or 3 standard deviations. See text for more details.}
{Note: the}  experimental sensitivity is expected to be dominated by systematic uncertainties associated with the detector response, %and
parton shower and  fragmentation~modelling.
}
\end{figure}

%\section{Outlook at higher-energies}
\section{Different Energies and Colliders}\label{sec:higherenergies}

{Up to here,}
%Till now,
we {were} %have
focused on a future \eee collider operating at $\sqrt{s}=250$~GeV, but %likely
those machines
%will
{later may
run} at higher energies. %Therefore,
{To this end,} {we %have
extended}
%our
{the}
study to $\sqrt{s}=0.5$ and 1~TeV, at particle level for {now.} %MDPI: the paragraph rewritten as done for this paper, not beyond. Same just below. Please consider. %Author: accepted
%the time being.

As the HV signal $e^{+}e^{-}\to Q_v\bar{Q}_v \to \text{hadrons}$, we
{have considered} the two extreme cases: the lightest communicator
$D_v$ and the heaviest one $T_v$ which decays into $ q_v t$. %We have
{The findings confirm}  %confirmed
that intermediate masses yield intermediate results.
Besides the $q\bar{q}$ and \mbox{$WW\rightarrow 4q$} backgrounds, we also %take
{took} into account the production of $t\bar{t}$, which,
%in fact,
{actually}
becomes relevant at these energies.
Table~\ref{tab:cross_sections_higherenergy} presents the cross-sections obtained from \Pythia\ {simulation.}
%We set the
{The}
masses of $D_v$ and $T_v$ {were set to}
equal %to
$\sqrt{s}/2$, and results {obtained}
for different $m_{q_v}$ are shown {in Table \ref{tab:cross_sections_higherenergy}}. %MDPI: please confirm, otherwise not clear where. %Author: accepted
No %large
{significant} variations are {seen}
%obtained
around this mass assignment to the communicator. According to expectations, the contribution from the SM backgrounds
{is found to decrease}%s
with {the c.m.}
energy. For the HV signal, a reduction of the cross-section by two orders of magnitude is obtained at
\mbox{$\sqrt{s}=1$ TeV. }

\begin{table}[H]
\small
\caption{\label{tab:cross_sections_higherenergy} {Inclusive} %MDPI:  We moved Table after its first citation, please consider. %Author: accepted
cross-sections of the HV signal and SM background at $\sqrt{s}=500$ GeV and {\mbox{$%\sqrt{s}= %MDPI; no repetition necessary, clear enough. Please consider. %Author: accepted
1$ TeV}}. Cross-section predictions
%do not depend
{show no dependence} %MDPI: please confirm you tak about th eresults in this Table and not those known from beyond the paper.  %Author: accepted
on the $m_{q_v}$ value unless it is too large to make the process kinematically inaccessible.
The signal corresponds to $D_v$ and $T_v$ pair production with the $D_v$/$T_v$ {masses} %MDPI: plural, otherwise looks a ratio. Pleaae confirm not a ratio. %Author: confirm that it is not a ratio
set {to equal %to
$\sqrt{s}/2$.} Whenever kinematically
allowed, $t\bar{t}$ production has been included as a SM background source in addition to lighter flavours considered at $\sqrt{s}=250$ GeV.}
\begin{tabularx}{\textwidth}{CCCC}
\toprule
{\textbf{Model}} & %MDPI; we added this column which clarfies the model considered. Please confirm.  %Author: confirm
\textbf{Process} &  \boldmath{$\sigma_{\sqrt{s}~=~500~\text{\textbf{GeV}}}$}  {\bf {(}%[
pb{)}%]
}  & \boldmath{$\sigma_{\sqrt{s}~=~1~\text{\textbf{TeV}}}$} {\bf {(}%[
pb{)}%]
}   \\
\midrule
%& $m_{D_v}= 250~\gev$ & $m_{D_v}= 500~\gev$ \\
{HV} & $e^{+}e^{-}\to D_v\bar{D}_v $ & $2.4\times 10^{-2}~\break (m_{D_v}= 250~\gev)$ %MDPI: moved to the next line in the same column. Same for other occasions. Please consider. %Author: accepted
& $4.4\times 10^{-3}~\break (m_{D_v}= 500~\gev)$
\\
%& $m_{T_v}= 250$ GeV  & $m_{T_v}= 500$ GeV  \\
&
$e^{+}e^{-}\to T_v\bar{T}_v$  & $9.5 \times 10^{-2}~\break (m_{T_v}~= 250~\gev)$& $1.8\times 10^{-2}~\break (m_{T_v}~= 500~\gev)$ \\
\midrule
{SM} & $e^{+}e^{-}\to q\bar{q}$ with ISR & 11 & 2.9  \\
& $e^{+}e^{-}\to t\bar{t}$  & $0.59$ & $0.19$ \\
& $WW\rightarrow4q$ & 3.4 & 1.3 \\
\bottomrule
\end{tabularx}
\end{table}

This study is also relevant for the \eee option of the Future Circular Collider (FCC) operating at %an
{c.m.} energy of
$\sqrt{s}=240~\gev$, i.e., {%very near
close} to the one {mainly considered here.} %MDPI: fits better. Please consider.  %Author: accepted
%assumed so far.
{\color{black} Studies with a preliminary version of the ILD
concept adapted to FCC have been started~\cite{fcc-note}. The limited coverage in the forward region makes ISR-photon detection
less efficient; however, the pertinent optimisation of selection cuts recovers sensitivity to an HV signal~\cite{fcc-note}.
Optimization of the ILD design for adaptation to FCC conditions is also to be performed. Similar prospects should be expected for
the Chinese Circular Electron Positron Collider (CepC), which shares several design features with the FCC. All Higgs %F
{factories}
concepts (linear or circular) share {quite} %very
ambitious designs for their detectors, featuring {high-quality}
%superb %MDPI: colloquial, not recommended for scientific papers. Please consider. %Author: accepted
tracking and vertexing capabilities, including {quite} large acceptance, as discussed briefly in Section~\ref{sec:sens}.
{Thus, the} method proposed in this
study {is %thus
applicable} across various collider~designs.}

\section{Conclusions}
\label{sec:conclusions}
The {study} %analysis
of particle correlations in high-energy colliders  %can provide
{provides} valuable insights into matter under extreme temperature and density
conditions, somehow reproducing the early-universe conditions when quarks and gluons had not yet bound to form hadrons.
On the other hand, this kind of {investigation} %analysis
%can
{may}
become a complementary tool to other conventional searches to uncover the existence of new phenomena including BSM physics
%beyond the SM, as {argued} %postulated
in Refs.~\cite{Sanchis-Lozano:2008zjj,Sanchis2009} and {have been} studied in this {paper}.
%work.

Motivated by the experimental observation of unexpected structures shown in angular correlations from hadronic collisions, we have
explored at detector level the discovery potential for hidden sectors at future \eee colliders using two-particle angular
correlations. Specifically, we {have considered}
a QCD-like HV model containing {not excessively heavy} %not-too-heavy
$v$-quarks and $v$-gluons, which {may} %MDPI: or "considered to" %can  %Author: accepted
interact with the SM sector via communicators of mass, typically {below about} %$\lesssim$
1~\tev. We have focused on $D_v\bar{D}_v$ pair production at
$\sqrt{s}=250$ GeV, of the order of the expected mass for the lightest communicator in this scenario. We %have
briefly extended our
study {here} at the particle level to higher energies (up to 1~TeV), {which pointed at} %pointing out
promising {prospects.} % too. %MDPI: rewritten, fits better. Please confirm the intended meaning have been retained. %Author: accepted

To conclude, %our
{the
results obtained here}
show that two-particle azimuthal correlations in a \eee Higgs factory could indeed become a useful tool
for discovering NP {as} %if it is
kinematically accessible. Although a specific
HV model has been employed {in this tudy}, other types of hidden sectors %would expectedly
{believed to} yield similar signatures.
In addition, collective effects stemming from a source different from BSM cannot be excluded in this kind of
%analysis.
{studies.} %MDPI: and plural. %Author: accepted
Such searches, based on rather diffuse signals that spread over
a large {enough} number of final-state particles, {to} %should
be contemplated as
complementary to other more conventional methods, thereby increasing the discovery potential of %these machines.
{Higgs factories.}

%\clearpage
\vspace{6pt}
\authorcontributions{{The authors contributed equally to this work. }
All authors have read and agreed to the published version of the manuscript. %For research articles with several· authors, a short paragraph specifying their individual contributions must be provided. The following statements should be used ``Conceptualization, X.X. and Y.Y.; methodology, X.X.; software, X.X.; validation, X.X., Y.Y. and Z.Z.; formal analysis, X.X.; investigation, X.X.; resources, X.X.; data curation, X.X.; writing---original draft preparation, X.X.; writing---review and editing, X.X.; visualization, X.X.; supervision, X.X.; project administration, X.X.; funding acquisition, Y.Y. All authors have read and agreed to the published version of the manuscript.'', please turn to the  \href{http://img.mdpi.org/data/contributor-role-instruction.pdf}{CRediT taxonomy} for the term explanation. Authorship must be limited to those who have contributed substantially to the work~reported. Otherwise "The authors contributed equally to this work".
%Author, added
}

\funding{A.I., E.M. and V.A.M. acknowledge {support} %MDPI: the financial information moved to  the Funding section form  the  Acknowledgments.Please consider. %Author: accepted
by Spanish MCIU {(Ministerio de Ciencia, Innovaci\'on y Universidades)}%MDPI: No undefined abbreviations. Please confirm thsi and the following abbreviations. %Author: confirm
/AEI {(Agencia Estatal de Investigaci\'on)}/10.13039/501100011033  and European Union {(EU)}/FEDER {(Fondo Europeo de
Desarrollo Regional)} via the grant PID2021-122134NB-C21.
E.M. and V.A.M. acknowledge support by Generalitat Valenciana (GV){, Spain,}  via the Excellence Grant CIPROM/2021/073.  %Author: this is the wording that they require us to use.
A.I. acknowledges support by GV under the grant number {CIDEGENT}%MDPI: Please provide the full-wording. %Author: this is the wording that they require us to use.
/2020/21, the financial support from the {MCIU} %MDPI: shoudn't this ne "MCIU"? If yes, please revise. If not, please provide the full-wording.   %Author: MCIU yes
with funding from the European Union NextGenerationEU and by GV via the Programa de Planes Complementarios de I+D+i (PRTR 2022) Project \textit{Si4HiggsFactories},
reference {ASFAE}%MDPI: Please provide the full-wording.  %Author: this is the wording that they require us to use.
$/2022/015$.
I.C. acknowledges support by the Xunta de Galicia (CIGUS {(Centros de investigaci\'on del Sistema Universitario de Galicia)} Network of
Research Centers).
M.\'A.S.-L. acknowledges support from the Spanish Agency  Estatal de Investigacion under Grant PID2023-151418NB-I00 funded by
MCIU/AEI/10.13039/501100011033/ FEDER, {EU} %MDPI: Shoudn't this be "EU"? Please revise as so. If not, please provide the full-wording. Confirmed, it is EU (the confusion came from the spanish version of it, which is inverted)
and by GV under grant CIPROM/2022/36.
A.I., E.M., V.A.M. and M.\'A.S.-L. acknowledge support by the Spanish MCIU/AEI via the Severo Ochoa project CEX2023-001292-S.
}

\dataavailability{Simulations and software repositories can be shared under reasonable~requests.%We encourage all authors of articles published in MDPI journals to share their research data. In this section, please provide details regarding where data supporting reported results can be found, including links to publicly archived datasets analyzed or generated during the study. Where no new data were created, or where data is unavailable due to privacy or ethical restrictions, a statement is still required. Suggested Data Availability Statements are available in section ``MDPI Research Data Policies'' at \url{https://www.mdpi.com/ethics}.
}

\acknowledgments{We thank {Mikael}%MDPI: it is highly recommende and always nice to provide also the first name(s) of th eoperson been acknowledged.Please consider. %Author: done
~Berggren and the ILD software working group for their support with the usage of detector fast simulation and the reconstruction tools employed in this study.
} %Author, added Mikael

\conflictsofinterest{The authors declare no conflicts of interest.%Declare conflicts of interest or state ``The authors declare no conflicts of interest.'' Authors must identify and declare any personal circumstances or interest that may be perceived as inappropriately influencing the representation or interpretation of reported research results. Any role of the funders in the design of the study; in the collection, analyses or interpretation of data; in the writing of the manuscript; or in the decision to publish the results must be declared in this section. If there is no role, please state ``The funders had no role in the design of the study; in the collection, analyses, or interpretation of data; in the writing of the manuscript; or in the decision to publish the results''.
}  %Author, added

\abbreviations{{Abbreviations}}{\vspace{-7pt}

%MDPI: many abbreviations used, some not even defined, so listed here to help the readers even given in the text.  No undefined abbreviations, variables, quantities, indexes etc. as met first in each of the main parts of a paper even known. Please consider. Please add if any is missing. %Author: revised. Minor chenges implemented and indicated
\noindent The following abbreviations are used in this manuscript:\\

\noindent
\hspace{-6pt}\begin{tabular}{p{1.5cm}p{11cm}}
ALEPH & Apparatus for LEP PHysics (detector, experiment  and Collaboration)  \\ %Author, corrected
Belle & name (experiment and Collaboration) \\
BSM & beyond the SM \\
BNL & Brookhaven National Laboratory \\
B1 & selection criterion (see text)\\
CERN & European Organization for Nuclear Research \\
CepC & Circular Electron Positron Collider \\
CGC & color glass condensate \\
c.m. & center-of-mass \\
cont. & continuum (see text) \\
\multirow{2}{*}{\textsc{Herwig}} & Hadron Emission Reactions {W}ith Interfering Gluons (high-energy physics Monte  \\
& {Carlo}  {generator)}\\
\end{tabular}
}
\abbreviations{}{%\vspace{-7pt}

%MDPI: many abbreviations used, some not even defined, so listed here to help the readers even given in the text.  No undefined abbreviations, variables, quantities, indexes etc. as met first in each of the main parts of a paper even known. Please consider. Please add if any is missing. %Author: revised. Minor chenges implemented and indicated
%\noindent The following abbreviations are used in this manuscript:\\

\noindent
\hspace{-6pt}\begin{tabular}{p{1.5cm}p{11cm}}

FCC & Future Circular Collider \\
HV & Hidden valley \\
H20 & name (ILC program of single-Higgs measurements) \\
ISR & initial state radiation \\
ILC & International Linear Collider \\
ILD & International Large Detector \\
{inv} & invariant \\
LHC & Large Hadron Collider \\  %Author, corrected (capital L)
NP & new physics \\
PFO & particle flow object\\
\textsc{Pythia} & name (high-energy physics Monte Carlo generator) \\
QCD &  quantum chromodynamics \\
RHIC &  Relativistic Heavy Ion Collider \\
RMS & root mean square \\
RR & radiative return \\
SGV & Simulation \`a Grande Vitesse \\
SM & Standard Model \\
S1, S2, S3 & selection criteria (see text)
\end{tabular}
}

%\bibliographystyle{JHEP}

%\bibliography{references}

\begin{thebibliography}{999}
\bibitem[Kittel and De~Wolf(2005)]{Kittel:2005fu}
Kittel, W.; De~Wolf, E.A.
\newblock {\em {Soft Multihadron Dynamics}}; World Scientific Publishing Co. Pte. Ltd.:
{Singapore,} 
2005.
\newblock {. [\href{http://doi.org/10.1142/5805}{CrossRef}]

\bibitem[Botet and Ploszajczak(2002)]{Botet:2002gj}
Botet, R.; Ploszajczak, M.
\newblock {\em {Universal Fluctuations: The Phenomenology of Hadronic Matter}};
World Scientific Publishing Co. Pte. Ltd.: {Singapore,}  2002. [\href{http://dx.doi.org/10.1142/4916}{CrossRef}]

\bibitem[Adams et~al.(2005)]{STAR:2005ryu}
Adams, J.  {et~al. [STAR Collaboration]} 
\newblock {Distributions of charged hadrons associated with high transverse
momentum particles in $pp$ and $\rm Au + Au$ collisions at $\sqrt{s_{NN}} = 200$~{GeV}}. 
\newblock {\em Phys. Rev. Lett.} {\bf 2005}, {\em 95},~152301. [\href{http://dx.doi.org/10.1103/PhysRevLett.95.152301}{CrossRef}] [\href{http://www.ncbi.nlm.nih.gov/pubmed/16241721}{PubMed}]

\bibitem[Alver et~al.(2010{\natexlab{a}})]{PHOBOS:2008yxa}
Alver, B.;  Back, B.B.; Baker, M.D.; Ballintijn, M.; Barton, D.S.; Betts, R.R.; Bindel, R.; Busza, W.; Chai, Z.; Chetluru, V.;
{et~al. 
}
\newblock {System size dependence of cluster properties from two-particle
angular correlations in $\rm Cu+Cu$ and $\rm Au + Au$ collisions at $\sqrt{s_{NN}}=
200$~GeV}. 
\newblock {\em Phys. Rev. C} {\bf 2010}, {\em 81},~024904. [\href{http://dx.doi.org/10.1103/PhysRevC.81.024904}{CrossRef}]

\bibitem[Alver et~al.(2010{\natexlab{b}})]{PHOBOS:2009sau}
Alver, B.;
Back, B.B.; Baker, M.D.; Ballintijn, M.; Barton, D.S.; Betts, R.R.; Bickley, A.A.;
Bindel, R.; Busza, W.; Carrol, A.;
{et~al. 
}
\newblock {High transverse momentum triggered correlations over a large
pseudorapidity acceptance in $\rm Au + Au$ collisions at  $\sqrt{s_{NN}}=
200~\text{GeV}$}.
\newblock {\em Phys. Rev. Lett.} {\bf 2010}, {\em 104},~062301. [\href{http://dx.doi.org/10.1103/PhysRevLett.104.062301}{CrossRef}] [\href{http://www.ncbi.nlm.nih.gov/pubmed/20366815}{PubMed}]

\bibitem[Abelev et~al.(2009)]{STAR:2009ngv}
Abelev, B.I.  {et~al. [STAR Collaboration]}
\newblock {Long range rapidity correlations and jet production in high energy
nuclear collisions}.
\newblock {\em Phys. Rev. C} {\bf 2009}, {\em 80},~064912. [\href{http://dx.doi.org/10.1103/PhysRevC.80.064912}{CrossRef}]

\bibitem[Aamodt et~al.(2010)]{ALICE:2010suc}
Aamodt, K.  {et~al. [ALICE Collaboration]}
\newblock {Elliptic flow of charged particles in Pb--Pb collisions at $\sqrt{s_{NN}}=2.76~\text{TeV}$}.
\newblock {\em Phys. Rev. Lett.} {\bf 2010}, {\em 105},~252302. [\href{http://dx.doi.org/10.1103/PhysRevLett.105.252302}{CrossRef}] [\href{http://www.ncbi.nlm.nih.gov/pubmed/21231580}{PubMed}]

\bibitem[Chatrchyan et~al.(2011)]{CMS:2011cqy}
Chatrchyan, S.  {et~al. [The CMS Collaboration]}
\newblock {Long-range and short-range dihadron angular correlations in central
PbPb collisions at $\sqrt{s_{\rm NN}}=$2.76 TeV}.
\newblock {\em J. High Energy Phys.} {\bf 2011}, {\em 2011
},~76. [\href{http://dx.doi.org/10.1007/JHEP07(2011)076}{CrossRef}]

\bibitem[Chatrchyan et~al.(2012)]{CMS:2012xss}
Chatrchyan, S.  {et~al. [The CMS Collaboration]}
\newblock {Centrality dependence of dihadron correlations and azimuthal
anisotropy harmonics in PbPb collisions at $\sqrt{s_{NN}}=2.76~\text{TeV}$}.
\newblock {\em Eur. Phys. J. C} {\bf 2012}, {\em 72},~2012. [\href{http://dx.doi.org/10.1140/epjc/s10052-012-2012-3}{CrossRef}]

\bibitem[Aad et~al.(2012)]{ATLAS:2012at}
Aad, G.  {et~al. [ATLAS Collaboration]}
\newblock {Measurement of the azimuthal anisotropy for charged particle
production in $\sqrt{s_{NN}}=2.76$ TeV lead--lead collisions with the ATLAS
detector}.
\newblock {\em Phys. Rev. C} {\bf 2012}, {\em 86},~014907. [\href{http://dx.doi.org/10.1103/PhysRevC.86.014907}{CrossRef}]

\bibitem[Chatrchyan et~al.(2014)]{CMS:2013bza}
Chatrchyan, S.  {et~al. [The CMS Collaboration]}
\newblock {Studies of azimuthal dihadron correlations in ultra-central PbPb
collisions at $\sqrt{s_{NN}} =$ 2.76 TeV}.
\newblock {\em J. High Energy Phys.} {\bf 2014}, {\em 2014
},~88. [\href{http://dx.doi.org/10.1007/JHEP02(2014)088}{CrossRef}]

\bibitem[Khachatryan et~al.(2010)]{CMS:2010ifv}
Khachatryan, V. {et~al. [The CMS Collaboration]}
\newblock {Observation of long-range, near-side angular correlations in
proton--proton collisions at the LHC}.
\newblock {\em J.~High Energy Phys.} {\bf 2010}, {\em 2010
},~91. [\href{http://dx.doi.org/10.1007/JHEP09(2010)091}{CrossRef}]

\bibitem[Chatrchyan et~al.(2013)]{CMS:2012qk}
Chatrchyan, S.  {et~al. [CMS Collaboration]}
\newblock {Observation of long-range, near-side angular correlations in
pPb collisions at the LHC}.
\newblock {\em Phys. Lett. B} {\bf 2013}, {\em 718},~795--814. [\href{http://dx.doi.org/10.1016/j.physletb.2012.11.025}{CrossRef}]

\bibitem[Abelev et~al.(2013)]{ALICE:2012eyl}
Abelev, B.  {et~al. [ALICE Collaboration]}
\newblock {Long-range angular correlations on the near and away side in p--Pb
collisions at $\sqrt{s_{\rm NN}}=5.02\text{TeV}$}.
\newblock {\em Phys. Lett. B} {\bf 2013}, {\em 719},~29--41. [\href{http://dx.doi.org/10.1016/j.physletb.2013.01.012}{CrossRef}]

\bibitem[Aad et~al.(2013)]{ATLAS:2012cix}
Aad, G.  {et~al. [ATLAS Collaboration]}
\newblock {Observation of associated near-side and away-side long-range
correlations in \mbox{$\sqrt{s_{NN}}=5.02~\text{TeV}$} proton--lead collisions with the
ATLAS detector}. 
\newblock {\em Phys. Rev. Lett.} {\bf 2013}, {\em 110},~182302. [\href{http://dx.doi.org/10.1103/PhysRevLett.110.182302}{CrossRef}] [\href{http://www.ncbi.nlm.nih.gov/pubmed/23683193}{PubMed}]

\bibitem[Chatrchyan et~al.(2013)]{CMS:2013jlh}
Chatrchyan, S.  {et~al. [CMS Collaboration]}
\newblock {Multiplicity and transverse momentum dependence of two- and
four-particle correlations in pPb and PbPb ollisions}.
\newblock {\em Phys. Lett. B} {\bf 2013}, {\em 724},~213--240. [\href{http://dx.doi.org/10.1016/j.physletb.2013.06.028}{CrossRef}]

\bibitem[Abelev et~al.(2013)]{ALICE:2013snk}
Abelev, B.B. {et~al. [ALICE Collaboration]}
\newblock {Long-range angular correlations of $\pi$, K and p in p--Pb
collisions at $\sqrt{s_{\rm NN}} = 5.02 \text{TeV}$}.
\newblock {\em Phys. Lett. B} {\bf 2013}, {\em 726},~164--177. [\href{http://dx.doi.org/10.1016/j.physletb.2013.08.024}{CrossRef}]

\bibitem[Khachatryan et~al.(2015{\natexlab{a}})]{CMS:2014und}
Khachatryan, V.  {et~al. [CMS Collaboration]}
\newblock {Long-range two-particle correlations of strange hadrons with charged
particles in pPb and PbPb collisions at LHC energies}.
\newblock {\em Phys. Lett. B} {\bf 2015}, {\em 742},~200--224. [\href{http://dx.doi.org/10.1016/j.physletb.2015.01.034}{CrossRef}]

\bibitem[Khachatryan et~al.(2015{\natexlab{b}})]{CMS:2015yux}
Khachatryan, V.  {et~al. [CMS Collaboration]} 
\newblock {Evidence for collective multiparticle correlations in $p$--Pb
collisions}.
\newblock {\em Phys. Rev. Lett.} {\bf 2015}, {\em 115},~012301. [\href{http://dx.doi.org/10.1103/PhysRevLett.115.012301}{CrossRef}] [\href{http://www.ncbi.nlm.nih.gov/pubmed/26182092}{PubMed}]

\bibitem[Aad et~al.(2016)]{ATLAS:2015hzw}
Aad, G.  {et~al. [ATLAS Collaboration]} 
\newblock {Observation of long-range elliptic azimuthal anisotropies in
$\sqrt{s}=$13 and 2.76 TeV $pp$ collisions with the ATLAS detector}.
\newblock {\em Phys. Rev. Lett.} {\bf 2016}, {\em 116},~172301. [\href{http://dx.doi.org/10.1103/PhysRevLett.116.172301}{CrossRef}] [\href{http://www.ncbi.nlm.nih.gov/pubmed/27176515}{PubMed}]

\bibitem[Khachatryan et~al.(2016)]{CMS:2015fgy}
Khachatryan, V.  {et~al. [CMS Collaboration]} 
\newblock {Measurement of long-range near-side two-particle angular
correlations in $pp$ collisions at $\sqrt s =13~\text{TeV}$}.
\newblock {\em Phys. Rev. Lett.} {\bf 2016}, {\em 116},~172302. [\href{http://dx.doi.org/10.1103/PhysRevLett.116.172302}{CrossRef}] [\href{http://www.ncbi.nlm.nih.gov/pubmed/27176516}{PubMed}]

\bibitem[Aaij et~al.(2016)]{LHCb:2015coe}
Aaij, R.  {et~al. [LHCb Collaboration]} 
\newblock {Measurements of long-range near-side angular correlations in
$\sqrt{s_{NN}}=5$~TeV proton--lead collisions in the forward region}.
\newblock {\em Phys. Lett. B} {\bf 2016}, {\em 762},~473--483. [\href{http://dx.doi.org/10.1016/j.physletb.2016.09.064}{CrossRef}]

\bibitem[Khachatryan et~al.(2017)]{CMS:2016fnw}
Khachatryan, V.  {et~al. [The CMS Collaboration]} 
\newblock {Evidence for collectivity in \emph{pp} collisions at the LHC}.
\newblock {\em Phys. Lett. B} {\bf 2017}, {\em 765},~193--220. [\href{http://dx.doi.org/10.1016/j.physletb.2016.12.009}{CrossRef}]

\bibitem[Aaboud et~al.(2017)]{ATLAS:2017hap}
Aaboud, M.  {et~al. [ATLAS Collaboration]} 
\newblock {Measurement of multi-particle azimuthal correlations in pp, $\rm p+Pb$
and low-multiplicity $\rm Pb+Pb$ collisions with the ATLAS detector}.
\newblock {\em Eur. Phys. J. C} {\bf 2017}, {\em 77},~428. [\href{http://dx.doi.org/10.1140/epjc/s10052-017-4988-1}{CrossRef}] [\href{http://www.ncbi.nlm.nih.gov/pubmed/29200942}{PubMed}]

\bibitem[Aaboud et~al.(2018)]{ATLAS:2017rtr}
Aaboud, M.  {et~al. [ATLAS Collaboration]} 
\newblock {Measurement of long-range multiparticle azimuthal correlations with
the subevent cumulant method in $pp$ and $p$+$Pb$ collisions with the ATLAS
detector at the CERN Large Hadron Collider}.
\newblock {\em Phys. Rev. C} {\bf 2018}, {\em 97},~024904. [\href{http://dx.doi.org/10.1103/PhysRevC.97.024904}{CrossRef}]

\bibitem[Aad et~al.(2023)]{ATLAS:2023bmp}
Aad, G. {et~al. [ATLAS Collaboration]} 
\newblock {Measurement of the sensitivity of two-particle correlations in $pp$
collisions to the presence of hard scatterings}.
\newblock {\em Phys. Rev. Lett.} {\bf 2023}, {\em 131},~162301. [\href{http://dx.doi.org/10.1103/PhysRevLett.131.162301}{CrossRef}] [\href{http://www.ncbi.nlm.nih.gov/pubmed/37925689}{PubMed}]

\bibitem[Hayrapetyan et~al.(2024)]{CMS:2023iam}
Hayrapetyan, A.  {et~al. [CMS Collaboration]} 
\newblock {Observation of enhanced long-range elliptic anisotropies inside
high-multiplicity jets in $pp$ collisions at $\sqrt{s}=13$~TeV}.
\newblock {\em Phys. Rev. Lett.} {\bf 2024}, {\em 133},~142301. [\href{http://dx.doi.org/10.1103/PhysRevLett.133.142301}{CrossRef}] [\href{http://www.ncbi.nlm.nih.gov/pubmed/39423390}{PubMed}]

\bibitem[CMS(2025)]{CMS:2025zhz}
{The CMS Collaboration.}
\emph{Unveiling the Dynamics of Long-Range Correlations in High-Multiplicity Jets
Through Substructure Engineering in pp Collisions at $\sqrt{s}=13$~TeV}; Report CMS PAS HIN-24-2024; CERN: Geneva, Switzerland, 2025.
\newblock{Available online: \url{https://cds.cern.ch/record/2930797} (accessed on 3 July 2025).} 


\bibitem[Shuryak(2007)]{Shuryak:2007fu}
Shuryak, E.V.
\newblock {
Origin of the 'ridge' phenomenon induced by jets in heavy ion
collisions}.
\newblock {\em Phys. Rev. C} {\bf 2007}, {\em 76},~047901. [\href{http://dx.doi.org/10.1103/PhysRevC.76.047901}{CrossRef}]

\bibitem[Dumitru et~al.(2008)Dumitru, Gelis, McLerran, and
Venugopalan]{Dumitru:2008wn}
Dumitru, A.; Gelis, F.; McLerran, L.; Venugopalan, R.
\newblock {Glasma flux tubes and the near side ridge phenomenon at RHIC}.
\newblock {\em Nucl. Phys. A} {\bf 2008}, {\em 810},~91--108. [\href{http://dx.doi.org/10.1016/j.nuclphysa.2008.06.012}{CrossRef}]

\bibitem[Bozek and Broniowski(2013)]{Bozek:2013uha}
Bo\.{z}ek, P.; Broniowski, W.
\newblock {Collective dynamics in high-energy proton--nucleus collisions}.
\newblock {\em Phys. Rev. C} {\bf 2013}, {\em 88},~014903. [\href{http://dx.doi.org/10.1103/PhysRevC.88.014903}{CrossRef}]

\bibitem[Dusling et~al.(2016)Dusling, Li, and Schenke]{Dusling:2015gta}
Dusling, K.; Li, W.; Schenke, B.
\newblock {Novel collective phenomena in high-energy proton--proton
and proton--nucleus collisions}.
\newblock {\em Int. J. Mod. Phys. E} {\bf 2016}, {\em 25},~1630002. [\href{http://dx.doi.org/10.1142/S0218301316300022}{CrossRef}]

\bibitem[Sanchis-Lozano and
Sarkisyan-Grinbaum(2017{\natexlab{a}})]{Sanchis-Lozano:2016qda}
Sanchis-Lozano, M.-A.; Sarkisyan-Grinbaum, E.
\newblock {A correlated-cluster model and the ridge phenomenon in
hadron--hadron collisions}.
\newblock {\em Phys. Lett. B} {\bf 2017}, {\em 766},~170--176. [\href{http://dx.doi.org/10.1016/j.physletb.2017.01.001}{CrossRef}]

\bibitem[Sanchis-Lozano and
Sarkisyan-Grinbaum(2017{\natexlab{b}})]{Sanchis-Lozano:2017aur}
Sanchis-Lozano, M.-A.; Sarkisyan-Grinbaum, E.
\newblock {Ridge effect and three-particle correlations}.
\newblock {\em Phys. Rev. D} {\bf 2017}, {\em 96},~074012. [\href{http://dx.doi.org/10.1103/PhysRevD.96.074012}{CrossRef}]

\bibitem[Noronha et~al.(2024)Noronha, Schenke, Shen, and Zhao]{Noronha:2024dtq}
Noronha, J.; Schenke, B.; Shen, C.; Zhao, W.
\newblock {Progress and challenges in small systems}.
\newblock
{\emph{Int. J. Mod. Phys. E} \textbf{2024}, {\it 33}, 2430005.}

\newblock [\href{http://dx.doi.org/10.1142/S0218301324300054}{CrossRef}]


\bibitem[Badea et~al.(2019)Badea, Baty, Chang, Innocenti, Maggi, Mcginn,
Peters, Sheng, Thaler, and Lee]{Badea:2019vey}
Badea, A.; Baty, A.; Chang, P.; Innocenti, G.M.; Maggi, M.; Mcginn, C.; Peters,
M.; Sheng, T.-A.; Thaler, J.; Lee, Y.-J.
\newblock {Measurements of two-particle correlations in $e^+e^-$ collisions at
91 GeV with ALEPH archived data}.
\newblock {\em Phys. Rev. Lett.} {\bf 2019}, {\em 123},~212002. [\href{http://dx.doi.org/10.1103/PhysRevLett.123.212002}{CrossRef}] [\href{http://www.ncbi.nlm.nih.gov/pubmed/31809127}{PubMed}]

\bibitem[Chen et~al.(2022)]{Belle:2022fvl}
Chen, Y.-C.  {et~al. [Belle Collaboration]} 
\newblock {Measurement of two-particle correlations of hadrons in $e^+e^-$
collisions at Belle}.
\newblock {\em Phys. Rev. Lett.} {\bf 2022}, {\em 128},~142005. [\href{http://dx.doi.org/10.1103/PhysRevLett.128.142005}{CrossRef}] [\href{http://www.ncbi.nlm.nih.gov/pubmed/35476485}{PubMed}]

\bibitem[Chen et~al.(2023)]{Belle:2022ars}
Chen, Y.-C.;  {et~al.}  [The Belle Collaboration] 
\newblock {Two-particle angular correlations in $e^{+}e^{-}$ collisions to
hadronic final states in two reference coordinates at Belle}.
\newblock {\em J. High Energy Phys.} {\bf 2023}, {\em 2023},~171. [\href{http://dx.doi.org/10.1007/JHEP03(2023)171}{CrossRef}]

\bibitem[Chen et~al.(2024)]{Chen:2023njr}

{Chen, Y.-C; Chen, Y.; Badea, A.; Baty, A.; Innocenti, G.M.; Maggi, M.; McGinn, C.; Peters, M.; Sheng, T.-A.; Thaler, J.; Lee, Y.-J;} 
\newblock {Long-range near-side correlation in $e^+e^-$ collisions at
183--209 GeV with ALEPH archived data}.
\newblock {\em Phys. Lett. B} {\bf 2024}, {\em 856},~138957. [\href{http://dx.doi.org/10.1016/j.physletb.2024.138957}{CrossRef}]

\bibitem[Gelis et~al.(2010)Gelis, Iancu, Jalilian-Marian, and
Venugopalan]{Gelis:2010nm}
Gelis, F.; Iancu, E.; Jalilian-Marian, J.; Venugopalan, R.
\newblock {The color glass condensate}.
\newblock {\em Ann. Rev. Nucl. Part. Sci.} {\bf 2010}, {\em 60},~463--489. [\href{http://dx.doi.org/10.1146/annurev.nucl.010909.083629}{CrossRef}]

\bibitem[Strassler and Zurek(2007)]{Strassler:2006im}
Strassler, M.J.; Zurek, K.M.
\newblock {Echoes of a hidden valley at hadron colliders}.
\newblock {\em Phys. Lett. B} {\bf 2007}, {\em 651},~374--379. [\href{http://dx.doi.org/10.1016/j.physletb.2007.06.055}{CrossRef}]

\bibitem[P\'erez-Ramos et~al.(2022)P\'erez-Ramos, Sanchis-Lozano, and
Sarkisyan-Grinbaum]{Perez-Ramos:2021mvm}
P\'erez-Ramos, R.; Sanchis-Lozano, M.-A.; Sarkisyan-Grinbaum, E.K.
\newblock {Searching for hidden matter with long-range angular correlations at
$e^+e^-$ colliders}.
\newblock {\em Phys. Rev. D} {\bf 2022}, {\em 105},~053001. [\href{http://dx.doi.org/10.1103/PhysRevD.105.053001}{CrossRef}]

\bibitem[T.~Lagouri (2022)]{Lagouri:2022ier}
Lagouri, T.
\newblock{Review on Higgs Hidden--Dark Sector physics at High-Energy Colliders}.
\newblock{{\it Symmetry} \textbf{2022}, {\it 14}, 1299.}
[\href{http://dx.doi.org/10.3390/sym14071299}{CrossRef}]

\bibitem[Strassler(2008)]{strassler2008unparticlemodelsmassgaps}
Strassler, M.J.
\newblock Why unparticle models with mass gaps are examples of hidden valleys.
{\emph{arXiv} \textbf{2008},	 arXiv:0801.0629.} 
\newblock [\href{http://dx.doi.org/10.48550/arXiv.0801.0629}{CrossRef}]

\bibitem[Knapen et~al.(2017)Knapen, Griso, Papucci, and Robinson]{Knapen_2017}
Knapen, S.; Griso, S.P.; Papucci, M.; Robinson, D.J.
\newblock Triggering soft bombs at the LHC.
\newblock {\em J. High Energy Phys.} {\bf 2017}, {\em 2017}, 76. [\href{http://dx.doi.org/10.1007/JHEP08(2017)076}{CrossRef}]

\bibitem[Albouy et~al.(2022)Albouy, Barron, Beauchesne, Bernreuther, Bona,
Cazzaniga, Cesarotti, Cohen, de~Cosa, Curtin, Demiragli, Doglioni, Elliot,
DiPetrillo, Eble, Erice, Freer, Garcia-Bellido, Gemmell, Genest, di~Cortona,
Gustavino, Hemme, Holmes, Kar, Knapen, Kulkarni, Lavezzo, Lowette, Maier,
Mee, Mrenna, Nair, Niedziela, Papageorgakis, Parmar, Paus, Pedro, Peixoto,
Perloff, Plehn, Scherb, Schwaller, Shelton, Singh, Sinha, Sjöstrand,
Spourdalakis, Stolarski, Strassler, Usachov, Sierra, Verhaaren, and
Wang]{Albouy_2022}
Albouy, G.; Barron, J.; Beauchesne, H.; Bernreuther, E.; Bona, M.; Cazzaniga,
C.; Cesarotti, C.; Cohen, T.; de~Cosa, A.; Curtin, D.;  {et~al.}
\newblock Theory, phenomenology, and experimental avenues for dark showers: A
Snowmass 2021 report.
\newblock {\em  Eur. Phys. J. C} {\bf 2022}, {\em 82}, 1132. [\href{http://dx.doi.org/10.1140/epjc/s10052-022-11048-8}{CrossRef}]

\bibitem[Beauchesne et~al.(2019)Beauchesne, Bertuzzo, and Grilli~di
Cortona]{Beauchesne_2019}
Beauchesne, H.; Bertuzzo, E.; Grilli~di Cortona, G.
\newblock Dark matter in Hidden Valley models with stable and unstable light
dark mesons.
\newblock {\em J. High Energy Phys.} {\bf 2019}, {\em 2019}, 118. [\href{http://dx.doi.org/10.1007/JHEP04(2019)118}{CrossRef}]

\bibitem[Sj\"ostrand et~al.(2015)Sj\"ostrand, Ask, Christiansen, Corke, Desai,
Ilten, Mrenna, Prestel, Rasmussen, and Skands]{Sjostrand:2014zea}
Sj\"ostrand, T.; Ask, S.; Christiansen, J.R.; Corke, R.; Desai, N.; Ilten, P.;
Mrenna, S.; Prestel, S.; Rasmussen, C.O.; Skands, P.Z.
\newblock {An introduction to PYTHIA 8.2}.
\newblock {\em Comput. Phys. Commun.} {\bf 2015}, {\em 191},~159--177. [\href{http://dx.doi.org/10.1016/j.cpc.2015.01.024}{CrossRef}]

\bibitem[Berggren(2012)]{berggren2012sgv}
Berggren, M.
\newblock {SGV 3.0---A fast detector simulation}. {\emph{arXiv} \textbf{2012},	arXiv:1203.0217.}  
\newblock [\href{http://dx.doi.org/10.48550/arXiv.1203.0217}{CrossRef}]

\bibitem[Abramowicz et~al.(2020)]{ILDConceptGroup:2020sfq}
Abramowicz, H. {et~al. [The ILD Concept Group]}
\newblock {International Large Detector: Interim Design Report}. {\emph{arXiv} {\bf 2020}, 	arXiv:2003.01116.} 
\newblock [\href{http://dx.doi.org/10.48550/arXiv.2003.01116}{CrossRef}]

\bibitem[F.~Gaede (2006)]{Gaede:2006pj}
Gaede, F.
\newblock{Marlin and LCCD---Software tools for the ILC}.
\newblock{\it Nucl. Instrum. Meth. Phys. Res. A Acceler. Spectrom. Detect. Assoc. Equip.} \textbf{2006} {\it 559}, 177--180.}
\newblock [\href{http://dx.doi.org/10.1016/j.nima.2005.11.138}{CrossRef}]

\bibitem[M.~A.~Thomson (2009)]{Thomson:2009rp}
Thomson, M.A. Particle flow calorimetry and the PandoraPFA algorithm.
\newblock{{\it Nucl. Instrum. Instrum. Meth. Phys. Res. A Acceler. Spectrom. Detect. Assoc. Equip.} \textbf{2009}, {\it 611}, 25--40.}
\newblock [\href{http://dx.doi.org/10.1016/j.nima.2009.09.009}{CrossRef}]


\bibitem[Musumeci et~al.(2023)Musumeci, Perez-Ramos, Irles, Corredoira, Mitsou,
Sarkisyan-Grinbaum, and Sanchis-Lozano]{Musumeci:2023rzf}
Musumeci, E.; Perez-Ramos, R.; Irles, A.; Corredoira, I.; Mitsou, V.A.;
Sarkisyan-Grinbaum, E.; Sanchis-Lozano, M.A.
\newblock {Two-particle angular correlations in the search for new physics at
future $e^+e^-$ colliders}.
\newblock {In Proceedings of the International Workshop on Future Linear
Colliders---LCWS 2023
},  {Menlo Park, CA, USA, 15--19 May 2023;} 
Brau, J.E., Ed.; SLAC eConf C23-05-15.3, online. [\href{http://dx.doi.org/10.48550/arXiv.2307.14734}{CrossRef}]

\bibitem[Irles et~al.(2024)Irles, M\'arquez, P\"oschl, Richard, Saibel,
Yamamoto, and Yamatsu]{Irles:2024ipg}
Irles, A.; M\'arquez, J.P.; P\"oschl, R.; Richard, F.; Saibel, A.; Yamamoto,
H.; Yamatsu, N.
\newblock {Probing gauge-Higgs unification models at the ILC with
quark--antiquark forward--backward asymmetry at
center-of-mass energies above the $Z$ mass}.
\newblock {\em Eur. Phys. J. C} {\bf 2024}, {\em 84},~537. [\href{http://dx.doi.org/10.1140/epjc/s10052-024-12918-z}{CrossRef}]

\bibitem[Bambade et~al.(2019)]{Bambade:2019fyw}
Bambade, P.;
Barklow, T.; Behnke, T.; Berggen, M.; Brau, J.; Burrows, P.; Denisov, D.; Faus-Golfe, A.; Foster, B.; Fujii, K.;
{et~al.}
\newblock {The International Linear Collider: A Global Project}. {\emph{arXiv} {\bf 2019}, arXiv:1903.01629.} 
\newblock [\href{http://dx.doi.org/10.48550/arXiv.1903.01629}{CrossRef}] [\href{http://www.ncbi.nlm.nih.gov/pubmed/40242606}{PubMed}]

\bibitem[Bahr et~al.(2008)]{Bahr:2008pv}
B\"ahr, M.; Gieseke, S.; Gigg, M.A.; Grellscheid, D.; Hamilton, K.; Latunde-Dada, O.; Pl\"atzer, S.; Richardson, P.; Seymour, M.H.;  {et~al.}
\newblock {Herwig++ physics and manual}.
\newblock {\em Eur. Phys. J. C} {\bf 2008}, {\em 58},~639--707. [\href{http://dx.doi.org/10.1140/epjc/s10052-008-0798-9}{CrossRef}]

\bibitem[Bellm et~al.(2016)]{Bellm:2015jjp}
Bellm, J.;
Gieske, S.;  Grellscheid, D.;  Pl\"atzer, S.; Rauch, M.; Reuschle, C.; Richardson, P.; Schichtel, P.;  Seymour, M.H.;  Si\'odmok, A.;
{et~al.}
\newblock {Herwig 7.0/Herwig++ 3.0 release note}.
\newblock {\em Eur. Phys. J. C} {\bf 2016}, {\em 76},~196. [\href{http://dx.doi.org/10.1140/epjc/s10052-016-4018-8}{CrossRef}]

\bibitem[Musumeci et~al.(2025)Musumeci, Irles, Perez-Ramos, Corredoira,
Sarkisyan-Grinbaum, Mitsou, and Sanchis-Lozano]{fcc-note}
Musumeci, E.; Irles, A.; Perez-Ramos, R.; Corredoira, I.; Sarkisyan-Grinbaum,
E.; Mitsou, V.A.; Sanchis-Lozano, M.A.
\newblock \emph{{Hidden Sectors at FCC-ee with Two-Particle Angular Correlations}};
{A Contribution to the FCC Feasibility Study in
Preparation for the Third European Strategy for Particle Physics Update,
January, 2026}; CERN: Geneva, Switzerland,  2025. [\href{http://dx.doi.org/10.17181/338f3-2c304}{CrossRef}]

\bibitem[Sanchis-Lozano(2009)]{Sanchis-Lozano:2008zjj}
Sanchis-Lozano, M.-A.
\newblock {Prospects of searching for (un)particles from hidden sectors using
rapidity correlations in multiparticle production at the LHC}.
\newblock {\em Int. J. Mod. Phys. A} {\bf 2009}, {\em 24},~4529--4572. [\href{http://dx.doi.org/10.1142/S0217751X09045820}{CrossRef}]

\bibitem[Sanchis-Lozano and Sarkisyan-Grinbaum(2018)]{Sanchis2009}
Sanchis-Lozano, M.A.; Sarkisyan-Grinbaum, E.K.
\newblock {Searching for new physics with three-particle correlations in $pp$
collisions at the LHC}.
\newblock {\em Phys. Lett. B} {\bf 2018}, {\em 781},~505--509. [\href{http://dx.doi.org/10.1016/j.physletb.2018.04.033}{CrossRef}]

\end{thebibliography}

\begin{adjustwidth}{-\extralength/4}{0cm}

\reftitle{References}

\PublishersNote{}

\end{adjustwidth}

\end{document}